\documentclass[twocolumn,showpacs,preprintnumbers,amsmath,amssymb]{revtex4}

\usepackage{graphicx}
\usepackage{dcolumn}
\usepackage{bm}

\usepackage{amssymb}
\usepackage{amsmath}

\begin{document}


\title{Evolution of the universe in entropic cosmologies via different formulations}

\author{Nobuyoshi {\sc Komatsu}$^{1}$}  \altaffiliation{E-mail: komatsu@se.kanazawa-u.ac.jp} 
\author{Shigeo     {\sc Kimura}$^{2}$}

\affiliation{$^{1}$Department of Mechanical Systems Engineering, Kanazawa University, 
                          Kakuma-machi, Kanazawa, Ishikawa 920-1192, Japan \\
                $^{2}$The Institute of Nature and Environmental Technology, Kanazawa University, 
                          Kakuma-machi, Kanazawa, Ishikawa 920-1192, Japan}%
\date{\today}

\begin{abstract}
We study two types of entropic-force models in a homogeneous, isotropic, spatially flat, matter-dominated universe.
The first type is a `$\Lambda(t)$ type' similar to $\Lambda(t)$CDM (varying-lambda cold dark matter) models in which both the Friedmann equation and the acceleration equation include an extra driving term.
The second type is a `BV type' similar to bulk viscous models in which the acceleration equation includes an extra driving term whereas the Friedmann equation does not.
In order to examine the two types systematically, we consider an extended entropic-force model that includes a Hubble parameter ($H$) term and a constant term in entropic-force terms.  
The $H$ term is derived from a volume entropy whereas the constant term is derived from an entropy proportional to the square of an area entropy.  
Based on the extended entropic-force model, we examine four models obtained from combining the $H$ and constant terms with the $\Lambda(t)$ and BV types.  
The four models agree well with the observed supernova data and describe the background evolution of the late universe properly. 
However, the evolution of first-order density perturbations is different in each of the four models, especially for low redshift, assuming that an effective sound speed is negligible.
The $\Lambda(t)$ type is found to be consistent with the observed growth rate of clustering, in contrast with the BV type examined in this study. 
A unified formulation is proposed as well, in order to examine density perturbations of the two types systematically. 
\end{abstract}

\maketitle

\section{Introduction}

$\Lambda$CDM (lambda cold dark matter) models are the simplest cosmological model which can explain an accelerated expansion of the late universe \cite{PERL1998ab,Riess1998_2004,Riess2007SN1,WMAP2011,Planck2013}.
However, the standard $\Lambda$CDM model (which assumes a cosmological constant $\Lambda$ and an additional energy component called `dark energy') suffers from several theoretical problems, e.g., the cosmological constant problem, the cosmic coincidence problem, etc. \cite{Weinberg1989}.
In order to solve the problems, various cosmological models have been suggested, using alternative dark energy, modified gravity, etc. (see, e.g., Refs.\ \cite{Weinberg1,Roy1,Miao1,Bamba1,Sola_2013b} and references therein).

In those models, two types of cosmological models have been extensively examined in an effort to explain an accelerated expansion of the universe. 
The first type is related to $\Lambda (t)$CDM models which assume a variable cosmological term $\Lambda(t)$ \cite{Freese1-Fritzsch1,Overduin1,Sola_2002,Sola_2003,Sola_2004,Sola_2009,Sola_2011a,Sola_2011b,Sola_2013a,Sola_2013b,Sola_2013c,Sola_2014b,Lima_2014a}. 
In this model, a varying $\Lambda(t)$ (which corresponds to an extra driving term) is added to both the Friedmann equation and the Friedmann--Lema\^{i}tre acceleration equation, instead of the cosmological constant $\Lambda$. 
We call this the `$\Lambda(t)$ type' of cosmological model.
The second type is related to bulk viscous models (which assume a bulk viscosity of cosmological fluids) \cite{Davies3,Weinberg0,Murphy1,Barrow11,Barrow12,Lima101,Zimdahl1,Arbab1,Brevik1,Brevik2,Nojiri1,Meng1,Fabris1,Colistete1,Barrow21,Meng2,Avelino1,Hipo1,Avelino2,Piattella1,Meng3,Pujolas1,Odintsov1,Odintsov2,Odintsov3,Odintsov4}  
and CCDM models (which assume a creation of cold dark matter)  \cite{Lima-Others1996-2008,Lima2010,Lima2010b,Lima2011,Lima2012}.
In the bulk viscous and CCDM models, the Friedmann equation does not include an extra driving term because dissipation processes are assumed.
We call this the `BV (bulk viscous) type'.

Recently, Easson \textit{et al.} have proposed an entropic-force model as an alternative explanation for the accelerated expansion of the universe \cite{Easson1,Easson2}.
We expect that the entropic-force model \cite{Easson1,Easson2,Koivisto1,Koma4,Koma4bc,Koma5,Cai1_Cai2_Qiu1,Casadio1-Costa1,Basilakos1,Lepe1,Sola_2014a} is related to $\Lambda(t)$ and BV types.
In the entropic-force model, an extra driving term, i.e., an entropic-force term, is derived from the usually neglected surface terms on the horizon of the universe \cite{Easson1}.
The entropic-force term can explain the accelerated expansion, without introducing new fields and an exotic energy component of the universe such as dark energy.
Instead of dark energy, the entropic-force model assumes that the horizon of the universe has an entropy and a temperature due to the information holographically stored there \cite{Easson1}.
For example, the Bekenstein entropy (area entropy) and the Hawking temperature are used in the original entropic-force model \cite{Easson1}.
The obtained entropic-force term is usually added to both the Friedmann and acceleration equations \cite{Easson1,Easson2,Koivisto1}.
(Note that the entropic-force considered here is different from the idea that gravity itself is an entropic-force \cite{Padma1,Verlinde1}.)
Accordingly, the original entropic-force model is $\Lambda(t)$ type.
The present authors have proposed a modified entropic-force model \cite{Koma5} assuming a generalized black-hole entropy proportional to its volume, based on appropriate nonadditive generalizations \cite{Tsallis2012}.  
The obtained entropic-force terms behave as if they were an extra driving term for bulk viscous models.
Therefore, the present authors have assumed dissipation processes similar to the bulk viscous model \cite{Koma5}.
That is, the modified entropic-force model corresponds to the BV type.
In this way, entropic-force models proposed so far can be categorized into $\Lambda(t)$ and BV types, using different formulations.

In entropic-force models, entropic-force terms depend on the class of entropy.
For example, $H^{2}$ terms are derived from an area entropy \cite{Easson1}, 
whereas $H$ terms are derived from a volume entropy \cite{Koma5}, where $H$ is the Hubble parameter.
The entropic-force terms affect the background evolution of the universe.
Note that we do not discuss inflation of the early universe.
In fact, entropic-force models which include $H^{2}$ terms cannot describe a decelerating and accelerating universe \cite{Basilakos1,Sola_2013a,Koma4}.
On the other hand, a modified entropic-force model which includes $H$ terms can describe a decelerating and accelerating universe \cite{Koma5}.
Of course, the entropic-force terms affect density perturbations as well.
For example, Basilakos \textit{et al.} have recently shown that the original entropic-force model (which includes $H^{2}$ terms) does not describe cosmological fluctuations properly without the inclusion of a constant term \cite{Basilakos1}. 
Also, they have found that $\Lambda(t)$CDM models (similar to the original entropic-force model) are not consistent with the structure formation data \cite{Sola_2009}.
Furthermore, Li and Barrow have explained that bulk viscous models (which include $H$ terms) are difficult to reconcile with astronomical observations of structure formations \cite{Barrow21}.
The previous works suggest that it is necessary to consider not only an $H$ term but also a constant entropic-force term.
(Entropic-force models which include $H$ terms have been recently investigated by Basilakos and Sol\`{a} \cite{Sola_2014a}.)

The constant term has been examined in $\Lambda(t)$CDM \cite{Sola_2009,Sola_2011b,Sola_2013a,Sola_2013b,Basilakos1} and CCDM models \cite{Lima-Others1996-2008,Lima2010,Lima2010b,Lima2011,Lima2012}, 
whereas $H$ terms have been investigated in $\Lambda(t)$CDM \cite{Sola_2009} and bulk viscous models \cite{Davies3,Weinberg0,Murphy1,Barrow11,Barrow12,Lima101,Zimdahl1,Arbab1,Brevik1,Brevik2,Nojiri1,Meng1,Fabris1,Colistete1,Barrow21,Meng2,Avelino1,Hipo1,Avelino2,Piattella1,Meng3,Pujolas1,Odintsov1,Odintsov2,Odintsov3,Odintsov4}.
In those works, the influence of the extra driving terms is focused on, and therefore, the difference between the $\Lambda(t)$ and BV types has not yet been discussed systematically. 
This is because the two types are usually categorized into different models, e.g., $\Lambda(t)$CDM, CCDM, and bulk viscous models.
(Background evolutions of the universe in the $\Lambda (t)$CDM and CCDM models have been discussed in Ref.\ \cite{Lima_2014a}.)
However, it is possible to examine the two types systematically through entropic-force models. 
Density perturbations of the two types are expected to be different from each other, because each continuity equation of cosmological fluids is different even if the background evolution of the universe is the same.
Therefore, in the present study, we examine the properties of the $\Lambda(t)$ and BV types of entropic-force models.
To this end, we consider an extended entropic-force model which includes $H$ and constant terms.
The constant term is derived from an entropy proportional to the square of an area entropy \cite{Koma5}.

The remainder of the present paper is organized as follows.
In Sec.\ \ref{General Friedmann equations and entropic-force models}, we present the general Friedmann and continuity equations and discuss entropic-force models.
In Sec.\ \ref{Entropic-force models}, we present a brief review of two types of standard entropic-force model, i.e., the $\Lambda (t)$ and BV types. 
In Sec.\ \ref{An extended entropic-force model}, in order to examine the two types systematically, we consider an extended entropic-force model which includes $H$ and constant terms. 
In Sec.\ \ref{Density perturbations}, we briefly review the density perturbations of the $\Lambda(t)$ and BV types, in the linear approximation. 
In addition, we discuss a unified formulation, in order to examine the density perturbations of the two types systematically. 
In Sec.\ \ref{The present models}, based on the extended entropic-force model, we propose four models obtained from combining the $H$ and constant terms with the $\Lambda(t)$ and BV types.
In Sec.\ \ref{Results}, we examine the evolution of the universe in the four entropic-force models. 
Finally, in Sec.\ \ref{Conclusions}, we present our conclusions.

\section{General Friedmann equations and entropic-force models} 
\label{General Friedmann equations and entropic-force models}

In the present paper, we consider a homogeneous, isotropic, and spatially flat universe 
and examine the scale factor $a(t)$ at time $t$ in the Friedmann--Lema\^{i}tre--Robertson--Walker metric \cite{Koma4,Koma4bc,Koma5}. 
The general Friedmann equation is given as 
\begin{equation}
  \left(  \frac{ \dot{a}(t) }{ a(t) } \right)^2  =  H(t)^2     =  \frac{ 8\pi G }{ 3 } \rho (t) + f(t) ,  
\label{eq:General_FRW01_f}
\end{equation}
and the general acceleration equation is  

\begin{equation}
  \frac{ \ddot{a}(t) }{ a(t) }   =  \dot{H}(t) + H(t)^{2}   
                                          =  -  \frac{ 4\pi G }{ 3 } \left ( \rho (t) + \frac{3p(t)}{c^2}  \right ) + g(t) ,
\label{eq:General_FRW02_g}
\end{equation}
where $H(t)$ is defined by
\begin{equation}
   H(t) \equiv   \frac{ da/dt }{a(t)} =   \frac{ \dot{a}(t) } {a(t)}  .
\label{eq:Hubble}
\end{equation}
$G$, $c$, $\rho(t)$, and $p(t)$ are the gravitational constant, the speed of light, the mass density of cosmological fluids, and the pressure of cosmological fluids, respectively.
$f(t)$ and $g(t)$ are general functions corresponding to extra driving terms discussed later.
We can obtain the general continuity equation from the general Friedmann and acceleration equations, because two of the three equations are independent.
From Eqs.\ (\ref{eq:General_FRW01_f}) and (\ref{eq:General_FRW02_g}), the general continuity equation \cite{Koma4} is given by
\begin{equation}
       \dot{\rho} + 3  \frac{\dot{a}}{a} \left (  \rho + \frac{p}{c^2}  \right )
          =  \frac{3}{4 \pi G} H \left(  - f(t) -  \frac{\dot{f}(t)}{2 H }  +  g(t)      \right )     .
\label{eq:drho_General0}
\end{equation}
For $\Lambda$CDM models \cite{Weinberg1,Roy1}, both $f(t)$ and $g(t)$ are set to be $\Lambda /3 $, where $\Lambda$ is a cosmological constant.
Therefore, the right-hand side of Eq.\ (\ref{eq:drho_General0}) is $0$ \cite{C01}.
However, in general, the right-hand side of Eq.\ (\ref{eq:drho_General0}) is non-zero.

For example, a non-zero right-hand side of the general continuity equation appears in $\Lambda (t)$CDM models \cite{Freese1-Fritzsch1,Overduin1,Sola_2002,Sola_2003,Sola_2004,Sola_2009,Sola_2011a,Sola_2011b,Sola_2013a,Sola_2013b,Sola_2013c,Sola_2014b,Lima_2014a} in which $f(t)=g(t)$ is assumed.  
The original entropic-force model suggested by Easson \textit{et al.} \cite{Easson1,Easson2} is similar to $\Lambda (t)$CDM models, as discussed later.
Substituting $f(t) = g(t)$ into Eq.\ (\ref{eq:drho_General0}), we have 
\begin{equation}
       \dot{\rho} + 3  \frac{\dot{a}}{a} \left (  \rho + \frac{p}{c^2}  \right )
           =  - \frac{3}{8 \pi G}  \dot{f}(t)     \quad  [\Lambda(t) \hspace{1mm} \rm{type}]  .
\label{eq:drho_General0_f=g}
\end{equation}
This is the general continuity equation for the $\Lambda(t)$ type. 
The right-hand side of Eq.\ (\ref{eq:drho_General0_f=g}) is not $0$ when ${f}(t)$ is not constant.
The $\Lambda(t)$ type can be interpreted as a kind of energy exchange cosmology in which the transfer of energy between two fluids is assumed \cite{Barrow22}, 
e.g., interacting quintessence \cite{Amendola1Zimdahl01}, the interaction between matter and radiation \cite{Davidson1Szy1}, the interaction between dark energy and dark matter \cite{Wang0102}, or the interaction between holographic dark energy and dark matter \cite{Pavon_2005}. 

As another example, a non-zero right-hand side of the general continuity equation appears in bulk viscous models, in which a bulk viscosity $\xi$ of cosmological fluids is assumed  
\cite{Davies3,Weinberg0,Murphy1,Barrow11,Barrow12,Lima101,Zimdahl1,Arbab1,Brevik1,Brevik2,Nojiri1,Meng1,Fabris1,Colistete1,Barrow21,Meng2,Avelino1,Hipo1,Avelino2,Piattella1,Meng3,Pujolas1,Odintsov1,Odintsov2,Odintsov3,Odintsov4}. 
(An effective pressure $p_{e}$, e.g., $ p_{e} (t) = p(t) - 3 \xi  H(t) $, is assumed as well.) 
Consequently, the Friedmann equation for the bulk viscous model does not include extra driving terms, i.e., $f(t)=0$.
The bulk viscosity is usually the only thing that can generate an entropy in a homogeneous and isotropic universe \cite{Lima101}.
The relationship between entropic-force and bulk viscosity has been discussed in Ref.\ \cite{Koma5}.
In fact, cosmological equations for a modified entropic-force model examined in Ref.\ \cite{Koma5} are similar to those for the bulk viscous model.
(As discussed later, similar cosmological equations are used in CCDM models \cite{Lima-Others1996-2008,Lima2010,Lima2010b,Lima2011,Lima2012}.) 
Substituting $f(t) = 0$ into Eq.\ (\ref{eq:drho_General0}), we have 
\begin{equation}
       \dot{\rho} + 3  \frac{\dot{a}}{a} \left (  \rho + \frac{p}{c^2}  \right )
          =  \frac{3}{4 \pi G} H  g(t)   \quad  [\rm{BV} \hspace{1mm} \rm{type}]      .
\label{eq:drho_General0_f=0}
\end{equation}
This is the general continuity equation for the BV type. 
We emphasize that the right-hand side of Eq.\ (\ref{eq:drho_General0_f=0}) is not $0$, even if $g(t)$ is constant, e.g., $g(t) = \Lambda/3$.
That is, Eq.\ (\ref{eq:drho_General0_f=0}) is essentially different from Eq.\ (\ref{eq:drho_General0_f=g}).
It is expected that the difference between Eqs.\ (\ref{eq:drho_General0_f=g}) and (\ref{eq:drho_General0_f=0}) affects the evolution of density perturbations, even if the background evolution of the universe is the same.

It is possible to consider two types of entropic-force models, namely, the $\Lambda(t)$ and BV types, and 
in Sec.\ \ref{Entropic-force models}, we review the two types of standard entropic-force model.
In Sec.\ \ref{An extended entropic-force model}, we discuss an extended entropic-force model which includes $H$ and constant entropic-force terms.
The derivation of the entropic-force terms is summarized in Appendix \ref{Derivation of entropic-force}.
We can derive $H^{2}$, $H$, and constant terms from an area entropy $S_{r2}$, a volume entropy $S_{r3}$, and an entropy $S_{r4}$ proportional to $r_{H}^{4}$, respectively, 
where $r_{H}$ is the Hubble horizon (radius) given by $c/H$.

Cosmological equations for the above discussed models are similar to those for entropic-force models examined in this study although the theoretical backgrounds are different.
We expect that the present study can help to investigate the cosmological models from different viewpoints.

\subsection{Two types of standard entropic-force model}
\label{Entropic-force models}

In entropic cosmology, the horizon of the universe is assumed to have an associated entropy and an approximate temperature due to the information holographically stored there \cite{Easson1,Easson2}. 
In Secs. \ref{f(t) =g(t)} and \ref{f(t) =0}, we present the $\Lambda(t)$ and BV types, respectively.
We note that entropic-force models discussed here are different from holographic dark energy models \cite{Pavon_2005}, although the holographic principle \cite{Hooft-Bousso} is applied to both models.

\subsubsection{$\Lambda(t)$ type  $[f(t) =g(t)]$} 
\label{f(t) =g(t)}

For the $\Lambda(t)$ type [$f(t)=g(t)$], we briefly review the original entropic-force model suggested by Easson \textit{et al.} \cite{Easson1,Easson2}.
Note that we neglect high-order terms for quantum corrections because we do not discuss the inflation of the early universe. 
In the original entropic-force model, entropic-force terms are summarized \cite{Koivisto1} as
\begin{equation}
  f(t)     =  \alpha_{1} H^2    + \alpha_{2} \dot{H} ,  
\label{eq:f_Koivisto}
\end{equation}
\begin{equation}
g(t)   =  \beta_{1} H^2  + \beta_{2} \dot{H}   , 
\label{eq:g_Koivisto}
\end{equation}
where $\alpha_1$ and $\alpha_2$ are expected to be given as 
\begin{equation}
\alpha_1 = \beta_1 \quad \textrm{and}  \quad \alpha_2 = \beta_2  .
\end{equation}
The four coefficients $\alpha_1$, $\alpha_2$, $\beta_1$, and $\beta_2$ are dimensionless constants.
In Refs.\ \cite{Easson2,Koivisto1}, it was argued that the extrinsic curvature at the surface was likely to result in 
something like 
\begin{equation}
   \alpha_1 = \beta_1 = 3/(2 \pi) \quad \textrm{and}  \quad   \alpha_2 = \beta_2 = 3/(4 \pi)    .
\end{equation}
In this way, general functions for the original entropic-force model are expected to be given by 
\begin{equation}
 f(t) = g(t) . 
\end{equation}
This type, i.e., the $\Lambda(t)$ type, has been the most typically examined entropic-force model \cite{Basilakos1}. 
The formulation can be interpreted as a modification of the left-hand side of the Einstein equation. 
The $H^{2}$ terms are derived from the Bekenstein entropy \cite{Easson1}, as shown in Appendix \ref{Entropic-force from the area entropy}.

$H^2$ and $\dot{H}$ terms included in Eqs.\ (\ref{eq:f_Koivisto}) and (\ref{eq:g_Koivisto}) have been investigated in $\Lambda(t)$CDM models \cite{Basilakos1,Sola_2013c}.
In those works, the influence of $\dot{H}$ terms was found to be similar to that of $H^2$ terms.
This implies that the $\dot{H}$ terms can be neglected \cite{Koma5}.
In fact, Easson {\it et al.} first proposed that the derived entropic-force terms are $H^2$ terms; i.e., $\dot{H}$ terms are not included in the entropic-force terms \cite{Easson1}. 
A similar fact was discussed in our previous works \cite{Koma4,Koma5}.
Therefore,  we neglect $\dot{H}$ terms, i.e., $\alpha_2 = \beta_2 =0$.

As examined in Refs.\ \cite{Basilakos1,Sola_2013a,Koma4,Koma5}, $H^2$ and $\dot{H}$ terms cannot describe a decelerating and accelerating universe predicted by the standard $\Lambda$CDM model. 
For example, Basilakos \textit{et al.} have shown it is not the $H^2$ and $\dot{H}$ terms, but rather an extra constant term that is important for describing a decelerating and accelerating universe \cite{Basilakos1}. 
The extra constant term for $\Lambda(t)$CDM models is naturally obtained from an integral constant of the renormalization group equation for the vacuum energy density \cite{Sola_2011b,Sola_2013b}.
A similar constant term appears in the creation of cold dark matter (CCDM) models \cite{Lima-Others1996-2008,Lima2010,Lima2010b,Lima2011,Lima2012}.
The CCDM model assumes a dissipation process based on gravitationally induced particle creation proposed by Prigogine \textit{et al.} \cite{Prigogine1989}.
That is, the CCDM model corresponds to the BV type, as discussed in the next subsection.

\subsubsection{BV type $[f(t) =0]$} 
\label{f(t) =0}

For the BV type [$f(t) =0$], we briefly review a modified entropic-force model based on an effective pressure \cite{Koma5}. 
The model discussed here is similar to both bulk viscous models \cite{Davies3,Weinberg0,Murphy1,Barrow11,Barrow12,Lima101,Zimdahl1,Arbab1,Brevik1,Brevik2,Nojiri1,Meng1,Fabris1,Colistete1,Barrow21,Meng2,Avelino1,Hipo1,Avelino2,Piattella1,Meng3,Pujolas1,Odintsov1,Odintsov2,Odintsov3,Odintsov4} 
and to the CCDM models \cite{Lima-Others1996-2008,Lima2010,Lima2010b,Lima2011,Lima2012}.
This is because an effective pressure is used in the two models, assuming dissipative processes.

In entropic-force models, $H^{2}$ terms are derived from the Bekenstein entropy (area entropy), 
whereas $H$ terms are derived from a generalized black-hole entropy (volume entropy) \cite{Koma5}. 
The $H^{2}$ and $H$ terms can be considered as extra driving terms.
Consequently, general functions are given by  
\begin{equation}
  f(t)     =  \alpha_{1} H^{2} + \hat{\alpha}_{3}  H, 
\label{eq:f_combined}
\end{equation}
\begin{equation}
g(t)   =  \beta_{1} H^{2} +  \hat{\beta}_{3} H  .
\label{eq:g_combined}
\end{equation}
When we assume an effective pressure \cite{Koma5}, $\alpha_{1}$ and $\hat{\alpha}_{3}$ are given as 
\begin{equation}
\alpha_{1} = 0  \quad     \textrm{and}    \quad  \hat{\alpha}_{3}  = 0 .
\end{equation}
$\hat{\alpha}_{3}$ and $\hat{\beta}_{3}$ are dimensional constants defined by  
\begin{equation}
  \hat{\alpha}_{3}   \equiv  \alpha_{3}  H_{0}      \quad     \textrm{and}    \quad     \hat{\beta}_{3} \equiv  \beta_{3} H_{0}   , 
\label{eq:ab3-H0_(2)}
\end{equation}
where $H_{0}$ is the Hubble parameter at the present time $t_{0}$.
The four coefficients $\alpha_1$, $\beta_1$, $\alpha_3$, and $\beta_3$ are dimensionless constants.
In this case, i.e., when we consider an effective pressure, the Friedmann equation does not include an extra driving term \cite{Koma5}.
(The formulation corresponds to a modification of the energy--momentum tensor of the Einstein equation.) 
Accordingly, general functions are summarized as
\begin{equation}
  f(t)     =  0 ,
\label{eq:f_combined2}
\end{equation}
\begin{equation}
g(t)   =  \beta_{1} H^{2} +  \hat{\beta}_{3} H  .
\label{eq:g_combined2}
\end{equation}
Because of the $H$ term, this model can predict a decelerating and accelerating universe, as in the case for a fine-tuned standard $\Lambda$CDM model \cite{Koma5}. 
However, as mentioned previously, similar cosmological models were found to be difficult to reconcile with astronomical observations of structure formations.  
Previous works imply that it is difficult to reconcile the standard entropic-force model with the structure formation data, without including constant terms.
However, we can derive a constant entropic-force term from an entropy $S_{r4}$ proportional to $r_{H}^{4}$, as shown in Appendix \ref{Entropic-force from the hyper-dimension entropy}. 
Therefore, in the next subsection, we consider an extended entropic-force model which includes the constant term.
Note that we assume $S_{r4}$ as one of the possible entropies.

\subsection{Extended entropic-force model}
\label{An extended entropic-force model}

In this subsection, we consider an extended entropic-force model which includes $H^{2}$, $H$, and constant entropic-force terms, in order to examine the $\Lambda(t)$ and BV types. 
Similar extra driving terms have been discussed in various cosmological models, 
e.g., $\Lambda(t)$CDM, CCDM, and bulk viscous models  \cite{C10}. 
However, their theoretical backgrounds are different from those of the entropic-force model.

As shown in Appendix \ref{Derivation of entropic-force}, the $H^{2}$, $H$, and constant terms are derived from an area entropy $S_{r2}$, a volume entropy $S_{r3}$, and an entropy $S_{r4}$ proportional to $r_{H}^{4}$, respectively. 
Using the three terms, the general functions are given by
\begin{equation}
  f(t) =  \alpha_{1} H^{2} + \hat{\alpha}_{3}  H   +  \hat{\alpha}_{4}   ,  
\label{eq:f(g)}
\end{equation}
\begin{equation}
  g(t) =  \beta_{1} H^{2} +  \hat{\beta}_{3} H   +  \hat{\beta}_{4}    , 
\label{eq:g(g)}
\end{equation}
where $\hat{\alpha}_{3}$, $\hat{\beta}_{3}$, $\hat{\alpha}_{4}$, and $\hat{\beta}_{4}$ are defined by  
\begin{equation}
                 \hat{\alpha}_{3}   \equiv  \alpha_{3}  H_{0},     \hspace{2mm}   \hat{\beta}_{3} \equiv  \beta_{3} H_{0},
\hspace{2mm}       \hat{\alpha}_{4}   \equiv  \alpha_{4}  H_{0}^{2},  \hspace{2mm}   \hat{\beta}_{4} \equiv  \beta_{4} H_{0}^{2}  .
\label{eq:ab3ab4-H0_(2)}
\end{equation}
The six coefficients $\alpha_1$, $\beta_1$, $\alpha_3$, $\beta_3$, $\alpha_4$, and $\beta_4$ are dimensionless constants.
The $\Lambda(t)$ and BV types are determined from the dimensionless coefficients.

We now consider the extended entropic-force model. 
From Eq.\ (\ref{eq:f(g)}), the modified Friedmann equation is written as 
\begin{equation}
  \left(  \frac{ \dot{a} }{ a } \right)^2   =  H^{2}  = \frac{ 8\pi G }{ 3 } \rho  + \alpha_{1} H^{2} + \hat{\alpha}_{3}  H   +  \hat{\alpha}_{4}   ,  
\label{eq:FRW01(g)}
\end{equation}
and, from Eq.\ (\ref{eq:g(g)}), the modified acceleration equation is written as 
\begin{align}
  \frac{ \ddot{a} }{ a } 
                     &=    \dot{H} +H^{2}                                                                                               \notag \\
                     &=    -  \frac{ 4\pi G }{ 3 } (  1+  3 w  ) \rho  +  \beta_{1} H^{2} +  \hat{\beta}_{3} H   +  \hat{\beta}_{4}    , 
\label{eq:FRW02(g)}
\end{align}
where $w$ is given by
\begin{equation}
  w = \frac{ p } { \rho  c^2 }       .
\label{eq:w_(2)}
\end{equation}
$w$ represents the equation of state parameter for a generic component of matter.
For non-relativistic matter (or a matter-dominated universe) $w$ is $0$, and for relativistic matter (or a radiation-dominated universe) $w$ is $1/3$.
In the present paper, we focus on a matter-dominated universe, i.e., $w=0$.
Note that, for generality, we leave $w$ in the following discussion.
Coupling [$(1+3w) \times $ Eq.\ (\ref{eq:FRW01(g)})] with [$2 \times $ Eq.\ (\ref{eq:FRW02(g)})] and rearranging, we obtain  
\begin{equation}
 \dot{H}  = \frac{ dH }{ dt }  = -  C_{1} H^2    +   \hat{C}_{3} H   +  \hat{C}_{4}, 
\label{eq:dHC1C3hC4h}
\end{equation}
where
\begin{equation}
  C_{1} = \frac{  3(1 + w)  -  \alpha_{1}(1+3w) - 2\beta_{1}  }{   2   } ,   
\label{eq:C1}
\end{equation}
\begin{equation}
  \hat{C}_{3} = \frac{  \hat{\alpha}_{3}(1+3w) + 2\hat{\beta}_{3}  }{   2   }  ,   
\hspace{2mm} \textrm{and} \hspace{2mm}   \hat{C}_{4} = \frac{  \hat{\alpha}_{4}(1+3w) + 2\hat{\beta}_{4}  }{   2   } .  
\label{eq:C3h_C4h}
\end{equation}
$C_{1}$ is a dimensionless parameter, whereas $\hat{C}_{3}$ and $\hat{C}_{4}$ are dimensional parameters.
Using Eq.\ (\ref{eq:ab3ab4-H0_(2)}), dimensionless parameters $C_{3}$ and $C_{4}$ are given by
\begin{equation}
  {C}_{3} = \frac{ \hat{C}_{3} }{ H_{0} } = \frac{ \alpha_{3}(1+3w) + 2 \beta_{3}  }{   2   } ,   
\label{eq:C3}
\end{equation}
\begin{equation}
  {C}_{4} = \frac{ \hat{C}_{4} }{ H_{0}^{2} } = \frac{ \alpha_{4}(1+3w) + 2 \beta_{4}  }{   2   }   .
\label{eq:C4}
\end{equation} 
The values of the six coefficients ($\alpha_1$, $\beta_1$, $\alpha_3$, $\beta_3$, $\alpha_4$, and $\beta_4$) for the $\Lambda(t)$ type are different from those for the BV type.
However, as shown in Sec. \ref{Results}, it is possible to determine two sets of six parameters under the condition that $C_{1}$, $C_{3}$, and $C_{4}$ for the $\Lambda(t)$ type are the same as those for the BV type. 
In this case, we can obtain the same solution from Eq.\ (\ref{eq:dHC1C3hC4h}).
That is, the background evolution of the universe of the BV type is equivalent to that of the $\Lambda(t)$ type. 
However, the continuity equation for the $\Lambda(t)$ type [Eq.\ (\ref{eq:drho_General0_f=g})] is different from Eq.\ (\ref{eq:drho_General0_f=0}) for the BV type. 
[Equation \ (\ref{eq:dHC1C3hC4h}) is essentially the same as a general $\Lambda(t)$CDM model examined by Basilakos \textit{et al.} \cite{Sola_2009}.
The solutions of Eq.\ (\ref{eq:dHC1C3hC4h}) are summarized in Appendix \ref{Solutions of an extended entropic-force model}.]

In Eqs.\ (\ref{eq:f(g)}) and (\ref{eq:g(g)}), the $H^{2}$ terms with $\alpha_1$ and $\beta_1$ are entropic-force terms.
The influence of the $H^2$ entropic-force terms is included in $C_{1}$ [Eq.\ (\ref{eq:C1})].
However, original $H^2$ terms cannot describe a decelerating and accelerating universe \cite{Koma5}.
Therefore, in this study, we neglect the $H^2$ entropic-force terms.
That is, $\alpha_1$ and $\beta_1$ are set to be $0$. 

As mentioned above, $H$ terms and constant terms ($C_{\rm{cst}}$ terms) are derived from a volume entropy and an entropy proportional to $r_{H}^{4}$, respectively. 
Accordingly, we consider the $H$ and constant $C_{\rm{cst}}$ terms separately.
We call them the `$H$ version' and `$C_{\rm{cst}}$ version', respectively. 
The $H$ and $C_{\rm{cst}}$ versions are discussed in Secs. \ref{H version} and \ref{Cst version}, respectively. 
In the following, we assume that $C_{1}$, $C_{3}$, and $C_{4}$ are positive constants.

\subsubsection{$H$ version} 
\label{H version} 

For the $H$ version, the general functions are given by
\begin{equation}
  f(t) =   \hat{\alpha}_{3}   H , 
\label{eq:f(H)}
\end{equation}
\begin{equation}
  g(t) =   \hat{\beta}_{3} H     .
\label{eq:g(H)}
\end{equation}
Consequently, Eq.\ (\ref{eq:dHC1C3hC4h}) is given as  
\begin{equation}
 \dot{H}    = -  C_{1} H^2    +   \hat{C}_{3}  H   , 
\label{eq:dHC1C3hC4h(H)}
\end{equation}
where $C_{1}$ and $\hat{C}_{3}$ are given by Eqs.\ (\ref{eq:C1}) and (\ref{eq:C3h_C4h}), respectively.
When $f(t)=0$, Eqs.\ (\ref{eq:f(H)})--(\ref{eq:dHC1C3hC4h(H)}) correspond to bulk viscous models studied in Refs.\ \cite{Davies3,Weinberg0,Murphy1,Barrow11,Barrow12,Lima101,Zimdahl1,Arbab1,Brevik1,Brevik2,Nojiri1,Meng1,Fabris1,Colistete1,Barrow21,Meng2,Avelino1,Hipo1,Avelino2,Piattella1,Meng3,Pujolas1,Odintsov1,Odintsov2,Odintsov3,Odintsov4}.  
The formulation of the $H$ version is essentially equivalent to that examined in our previous work \cite{Koma5}.
The evolution of the Hubble parameter is given as
\begin{equation}
 \frac{H} {H_{0}}  =  \left ( 1-  \frac{ C_{3} }{ C_{1} }  \right )   \left ( \frac{ a } {  a_{0} } \right )^{ -C_{1}}   +  \frac{ C_{3} }{ C_{1} }          ,
\label{eq:H/H0_(C1C3)_00}
\end{equation}
where $a_{0}$ represents the scale factor at the present time and $C_{3}$ is $\hat{C}_{3}/H_{0}$ given by Eq. (\ref{eq:C3}).

\subsubsection{$C_{\rm{cst}}$ version} 
\label{Cst version} 

For the $C_{\rm{cst}}$ version, the general functions are given by
\begin{equation}
  f(t) =   \hat{\alpha}_{4}    , 
\label{eq:f(Cst)}
\end{equation}
\begin{equation}
  g(t) =    \hat{\beta}_{4}     .
\label{eq:g(Cst)}
\end{equation}
Therefore, Eq.\ (\ref{eq:dHC1C3hC4h}) is given as  
\begin{equation}
 \dot{H}    = -  C_{1} H^2    +   \hat{C}_{4}    , 
\label{eq:dHC1C3hC4h(Cst)}
\end{equation}
where $C_{1}$ and $\hat{C}_{4}$ are given by Eqs.\ (\ref{eq:C1}) and (\ref{eq:C3h_C4h}), respectively.
Using $C_{4} =  \hat{C}_{4}/H_{0}^{2}$ [Eq. (\ref{eq:C4})], the evolution of the Hubble parameter is given as 
\begin{equation}
 \left ( \frac{H} {H_{0}} \right )^{2}   
=  \left ( 1-  \frac{ C_{4} }{ C_{1} }  \right )   \left ( \frac{ a } {  a_{0} } \right )^{ -2 C_{1}}   +  \frac{ C_{4} }{ C_{1} }     .
\label{eq:H/H0(C1C4)_00}
\end{equation}
When $f(t)=g(t)$, Eqs.\ (\ref{eq:f(Cst)})--(\ref{eq:H/H0(C1C4)_00}) are equivalent to those for the standard $\Lambda$CDM model.
The solution is summarized in Appendix \ref{Solutions for the model with constant terms}.

\section{First-order density perturbations}
\label{Density perturbations}

In the present paper, we examine density perturbations of the $\Lambda(t)$ and BV types, in the linear approximation.
To this end, we usually use two methods separately.
Accordingly, in Sec.\ \ref{Two methods for density perturbations}, we present the two methods. 
In Sec.\ \ref{Density perturbations for Lambda(t) types}, we review first-order density perturbations of the $\Lambda(t)$ type, according to the work of Basilakos \textit{et al.} \cite{Sola_2009}.
In Sec.\ \ref{Density perturbations for BV types}, we review first-order density perturbations of the BV type, according to the work of Jesus \textit{et al.} \cite{Lima2011}.
The two methods are expected to be summarized, using a neo-Newtonian approach proposed by Lima \textit{et al.} \cite{Lima_Newtonian_1997}.
Therefore, in Sec.\ \ref{Unified formulation}, we discuss a unified formulation based on the neo-Newtonian approach, in order to observe the density perturbations of the two types systematically.
In the present study, we focus on a matter-dominated universe.

\subsection{Two methods for density perturbations} 
\label{Two methods for density perturbations}

In this subsection, we present two methods to examine density perturbations of the $\Lambda(t)$ and BV types.
In Secs.\ \ref{Density perturbations for Lambda(t) types} and \ref{Density perturbations for BV types}, we review first-order density perturbations of the $\Lambda(t)$ and BV types, respectively.

\subsubsection{Density perturbations of the $\Lambda(t)$ type}
\label{Density perturbations for Lambda(t) types}

Density perturbations in $\Lambda(t)$CDM models have been closely examined, e.g., see the work of Basilakos \textit{et al.} \cite{Sola_2009}.
In fact, formulations of the $\Lambda(t)$ type discussed here are essentially equivalent to the $\Lambda(t)$CDM model.
Therefore, we only briefly review density perturbations of the $\Lambda(t)$ type, according to Ref.\ \cite{Sola_2009}.
To this end, we assume a matter-dominated universe ($w=0$), i.e., a pressureless fluid. 
Substituting $p=0$ into Eq.\ (\ref{eq:drho_General0_f=g}), the continuity equation for the $\Lambda(t)$ type becomes 
\begin{equation}
      \dot{\rho} + 3  \frac{\dot{a}}{a}   \rho   =  - \frac{3}{8 \pi G}  \dot{f}(t)   .
\label{eq:drho_General0_f=g_(p=0)}
\end{equation}
This equation is equivalent to the equation examined in Ref.\ \cite{Sola_2009}, when $8 \pi G = c \equiv 1$ and $3 \dot{f}(t)= \dot{\Lambda}(t)$.
(We consider the mass density for matter \cite{Waga1994}.) 
In Ref.\ \cite{Sola_2009}, Basilakos \textit{et al.} have focused on models where the time dependence of $\Lambda(t)$ appears always at the expense of an interaction with matter.
The model can be interpreted as an energy exchange cosmology, in which the transfer of energy between two fluids is assumed \cite{Barrow22}.
Similarly, in the present study, we assume an interchange of energy between the bulk (the universe) and the boundary (the horizon of the universe) \cite{Lepe1}, as if it were an energy exchange cosmology.
Consequently, the time evolution equation for the matter density contrast $\delta \equiv \delta \rho_{m} /\rho_{m} $, i.e., the perturbation growth factor, is given by \cite{Waga1994}
\begin{equation}
      \ddot{\delta} +  \left ( 2 H + Q \right ) \dot{\delta} - \left [ 4 \pi G \rho  - 2 H Q -\dot{Q} \right ] \delta = 0 ,
\label{eq:density_f=g}
\end{equation}
where 
\begin{equation}
 \rho =   \frac{3}{8 \pi G}  (H^{2} -f(t) ),     \quad   Q =  - \frac{3}{8 \pi G}  \frac{ \dot{f}(t) }{  \rho  }  .
\label{eq:Q_f=g}
\end{equation}
In the present paper, $\rho_{m}$ is replaced by $\rho$, because we consider a single-fluid-dominated universe.  
[Note that $\rho$ included in Eq.\ (\ref{eq:density_f=g}) represents $\bar{\rho}$ corresponding to a homogenous and isotropic solution for the unperturbed equations (the Friedmann, acceleration, and continuity equations).
For simplicity, $\bar{\rho}$ is replaced by $\rho$ when we present the time evolution equation for $\delta$.]
Substituting Eq.\ (\ref{eq:Q_f=g}) into Eq.\ (\ref{eq:drho_General0_f=g_(p=0)}), we obtain the continuity equation as
\begin{equation}
      \dot{\rho} + 3  \frac{\dot{a}}{a}   \rho   =   Q \rho  . 
\label{eq:drho__f=g_Q_(p=0)}
\end{equation}
When $8 \pi G = c \equiv 1$ and $3 \dot{f}(t)= \dot{\Lambda}(t)$, Eqs.\ (\ref{eq:Q_f=g}) and (\ref{eq:drho__f=g_Q_(p=0)}) are equivalent to those examined in Ref.\ \cite{Sola_2009}. 
Basilakos \textit{et al.} investigated density perturbations in various types of variable cosmological terms \cite{Sola_2009}. 
Therefore, we employ their theoretical solutions, as discussed  in Secs.\ \ref{Lambda(t)-H model} and \ref{Lambda(t)-Cst model}.
(In Sec.\ \ref{Unified formulation}, we discuss a unified formulation, in order to examine the $\Lambda(t)$ and BV types systematically.)

We note that it is necessary to define explicitly the functional form of the $f(t)$ component, in order to solve the above differential equation.
As described in Ref.\ \cite{Sola_2009}, the approach based on Eq.\ (\ref{eq:density_f=g}) implies that dark energy perturbations are negligible. 
This is justified in most cases \cite{Sola_2009,Sola_2007-2009}.
(For details, see Ref.\ \cite{Sola_2009}.) 
In the present study, we consider an interchange of energy between the bulk and the boundary.
Accordingly, we assume that boundary perturbations are negligible.

\subsubsection{Density perturbations of the BV type}
\label{Density perturbations for BV types}

The BV type discussed here is similar to both bulk viscous models \cite{Davies3,Weinberg0,Murphy1,Barrow11,Barrow12,Lima101,Zimdahl1,Arbab1,Brevik1,Brevik2,Nojiri1,Meng1,Fabris1,Colistete1,Barrow21,Meng2,Avelino1,Hipo1,Avelino2,Piattella1,Meng3,Pujolas1,Odintsov1,Odintsov2,Odintsov3,Odintsov4} 
and CCDM models \cite{Lima-Others1996-2008,Lima2010,Lima2010b,Lima2011,Lima2012}.
This is because these models assume dissipation processes and therefore an effective pressure must be employed.
In particular, density perturbations in the CCDM model have been closely examined, e.g., see the work of Jesus \textit{et al.} \cite{Lima2011}.
Accordingly, we review the  density perturbations of the BV type according to Ref.\ \cite{Lima2011}. 

In the present paper,  we assume a matter-dominated universe ($w=0$), i.e., a pressureless fluid. 
Substituting $p=0$ into Eq.\ (\ref{eq:drho_General0_f=0}), the general continuity equation for the BV type is given by 
\begin{equation}
      \dot{\rho} + 3  \frac{\dot{a}}{a}   \rho   =   \frac{3}{4 \pi G}  H g(t)   .
\label{eq:drho_General0_f=0_(p=0)}
\end{equation}
This equation is essentially equivalent to the equation examined in Ref.\ \cite{Lima2011}.
To confirm this, we consider an effective pressure $p_{e}$ due to dissipation processes.
The effective pressure $p_{e}$ is given by $p_{e}   =  p + p_{c} $,  
where $p_{c}$ is the creation pressure for constant specific entropy in the CCDM model \cite{Lima2011}.
In the present study, we interpret $p_{c}$ as a pressure derived from an entropic-force on the horizon of the universe.
Substituting $p=0$ into $p_{e}  =  p + p_{c} $,  we have $p_{e}  =  p_{c} $.
According to Ref.\ \cite{Lima2011}, $p_{c}$ for a CDM component can be expressed as 
\begin{equation}
p_{c} = - \frac{\rho c^2 \Gamma}{3H}    , 
\label{eq:pc_CCDM}
\end{equation}
where $\Gamma$ is given as 
\begin{equation}
\Gamma = \frac{3}{4 \pi G}  \frac{H g(t)}{ \rho }      .
\label{eq:gamma_CCDM}
\end{equation}
Therefore, Eq.\ (\ref{eq:drho_General0_f=0_(p=0)}) can be written as 
\begin{equation}
      \dot{\rho} + 3  \frac{\dot{a}}{a}   \rho   =   \Gamma \rho  .
\label{eq:drho_General0_f=0_gamma_(p=0)}
\end{equation}
Note that $\rho$ is the mass density for matter.
In the CCDM model, $\Gamma$ is the creation rate of CDM particles.
In the present study, we interpret $\Gamma$ as a parameter for entropy production processes.
From Eq.\ (\ref{eq:pc_CCDM}), the equation of state parameter is given by $ - \Gamma /(3H)$. 
Although Eq.\ (\ref{eq:drho_General0_f=0_gamma_(p=0)}) is similar to Eq.\ (\ref{eq:drho__f=g_Q_(p=0)}), $\Gamma$ is different from $Q$. 
(In Sec.\ \ref{Unified formulation}, we discuss a unified formulation based on a neo-Newtonian approach \cite{Lima2011,Lima_Newtonian_1997}, in order to examine the $\Lambda(t)$ and BV types systematically.) 

In general, a perturbation analysis in cosmology requires a full relativistic description \cite{Lima2011}. 
This is because the standard nonrelativistic (Newtonian) approach works well only when the scale of perturbation is much less than the Hubble radius and the velocity of peculiar motions is small in comparison to the Hubble flow. 
However, such difficulties are circumvented by the neo-Newtonian approximation, as described in Ref.\ \cite{Lima2011}. 
In fact, Jesus \textit{et al.} \cite{Lima2011} closely investigated density perturbations in the CCDM model corresponding to the BV type.
Therefore, we employ their formulations.
In our units, setting $c=1$, the time evolution equation for the matter density contrast $\delta$ is given by 
\begin{align}
\ddot{\delta}  & + \left [ 2 H + \Gamma + 3 c_{\rm{eff}}^{2} H  - \frac{ \Gamma \dot{H} - H \dot{\Gamma} }{ H (3H -\Gamma)}  \right ] \dot{\delta}        \notag \\
                     & +  \Bigg \{      3 (\dot{H} + 2 H^{2}) \left (   c_{\rm{eff}}^{2}  + \frac{ \Gamma }{ 3H } \right )     \notag \\
                     & + 3 H  \left [ \dot{c}_{\rm{eff}}^{2}   - ( 1 + c_{\rm{eff}}^{2}) \frac{ \Gamma \dot{H} - H \dot{\Gamma} }{ H (3H -\Gamma)}  \right ] \notag \\     
                     &  - 4 \pi G \rho \left ( 1 -  \frac{ \Gamma }{ 3H } \right ) ( 1 + 3 c_{\rm{eff}}^{2} )  + \frac{  k^{2} c_{\rm{eff}}^{2} }{ a^{2} }    \Bigg \}     \delta = 0              , 
\label{eq:delta-t_CCDM}
\end{align}
where an effective sound speed, $c_{\rm{eff}}^{2}$, is defined by 
\begin{equation}
 c_{\rm{eff}}^{2}  \equiv  \frac{\delta p_{c} }{\delta \rho }   . 
\label{eq:ceff2__CCDM}
\end{equation}
For simplicity, we set $c=1$ when we present the time evolution equation for $\delta$.
Jesus \textit{et al.} assumed  $c_{\rm{eff}}^{2} = c_{\rm{eff}}^{2}(t)$ and that the spatial dependence of $\delta$ is proportional to $e^{i \bf{k} \cdot \bf{x} }$, 
where the comoving coordinates $\bf{x}$ (which are related to the proper coordinates $\bf{r}$) are given by  $\bf{x} = \bf{r}$$/a$.
In Ref.\ \cite{Lima2011}, $c_{\rm{eff}}^{2}$ is considered to be a new degree of freedom, and the influence of $c_{\rm{eff}}^{2}$ on the density perturbations is examined in detail. 
In this paper, we consider $c_{\rm{eff}}^{2} = 0$. We explain the reason in Sec.\ \ref{BV-H model}.

In the present study, we assume a spatially flat universe, and therefore, the Friedmann equation is $4 \pi G \rho = 3 H^{2}/2 $.
In addition, for numerical purposes, we employ a new independent variable \cite{Lima2011} defined by 
\begin{equation}
\eta \equiv \ln (\tilde{a}(t)) ,  \quad \textrm{where} \quad  \tilde{a}(t) =  \frac{a(t)}{a_{0}}    .
\label{eta_def}
\end{equation}
Using these equations, Eq.\ (\ref{eq:delta-t_CCDM}) can be rearranged as 
\begin{equation}
\delta^{\prime \prime}  + F(\eta) \delta^{\prime}  +  G(\eta) \delta =0, 
\label{eq:delta-eta_c=0_CCDM_0}
\end{equation}
where $F(\eta)$ and $G(\eta)$ \cite{Lima2011} are given by
\begin{equation}
F(\eta) =  2  +  3 c_{\rm{eff}}^{2}  +  \frac{ \Gamma + H^{\prime} }{ H }   -   \frac{  \Gamma H^{\prime} - H \Gamma^{\prime}  }{ H (3H -\Gamma) }   ,
\label{eq:F(eta)_0}
\end{equation}
\begin{align}
G(\eta)  =  &   \left (  \frac{ \Gamma }{ H }  - 1   \right )   \left (  \frac{ \Gamma }{ 2H }  + \frac{3}{2}   + 3 c_{\rm{eff}}^{2}   \right )                    \notag \\   
                &    +   3 c_{\rm{eff}}^{2 \prime}   -  3 ( 1 + c_{\rm{eff}}^{2}  )  \frac{  \Gamma H^{\prime} - H \Gamma^{\prime}  }{ H (3H -\Gamma) }    + \frac{  k^{2} c_{\rm{eff}}^{2} e^{- 2 \eta}  }{ H^{2} }     .
\label{eq:G(eta)_0}
\end{align}
It should be noted that $^{\prime}$ represents the differential with respect to $\eta$, i.e., $d/d \eta$.
We can apply Eqs.\ (\ref{eq:delta-eta_c=0_CCDM_0})--(\ref{eq:G(eta)_0}) to various models for the BV type.
For example, the CCDM model proposed by Lima, Jesus, and Oliveira \cite{Lima2010} (the LJO model) was closely examined in Ref.\ \cite{Lima2011}.
Formulations of the LJO model are equivalent to those of a BV-$C_{\rm{cst}}$ model which includes a constant entropic-force term, as discussed in the next section.
In addition, we can apply Eqs.\ (\ref{eq:delta-eta_c=0_CCDM_0})--(\ref{eq:G(eta)_0}) to a BV-$H$ model which includes an $H$ term.
We discuss density perturbations in the BV-$H$ and BV-$C_{\rm{cst}}$ models in Secs. \ref{BV-H model} and \ref{BV-Cst model}, respectively.

The LJO model discussed above is obtained for a constant $g(t)$. 
In the LJO model \cite{Lima2011}, $\Gamma$ is defined by 
\begin{equation}
 \Gamma = 3 \tilde{\Omega}_{\Lambda} \left ( \frac{\rho_{c0}}{\rho} \right )  H  ,
\label{eq:gamma-H_cst_CCDM}
\end{equation}
where $\tilde{\Omega}_{\Lambda}$ is a constant parameter and $\rho_{c0}$ is the present value of the critical density given by $\rho_{c0} = 3 H_{0}^{2}/(8 \pi G)$.
In a spatially flat matter-dominated universe, the Hubble parameter is given by $H = H_{0} [ (1 - \tilde{\Omega}_{\Lambda}) \tilde{a}^{-3}  + \tilde{\Omega}_{\Lambda} ]^{1/2}$. 
This equation is equivalent to Eq.\ (\ref{eq:H/H0(C1C4)_00}) when $C_{1}=3/2$ and $\tilde{\Omega}_{\Lambda} = C_{4}/C_{1}$.

\subsection{Unified formulation for the $\Lambda(t)$ and BV types}
\label{Unified formulation}

In Secs.\ \ref{Density perturbations for Lambda(t) types} and \ref{Density perturbations for BV types}, we reviewed first-order density perturbations of the $\Lambda(t)$ and BV types separately. 
In this subsection, in order to examine entropic-force models systematically, we discuss a unified formulation for the $\Lambda (t)$ and BV types, using a neo-Newtonian approach.
The neo-Newtonian approach was proposed by Lima \textit{et al.} \cite{Lima_Newtonian_1997}, following earlier ideas developed by McCrea \cite{McCrea_1951} and Harrison \cite{Harrison_1965}, in order to describe a Newtonian universe with pressure \cite{Lima2011}. 
In fact, first-order density perturbations of the BV type discussed in Sec.\ \ref{Density perturbations for BV types} are derived from the neo-Newtonian approach.

As shown in Sec.\ \ref{General Friedmann equations and entropic-force models}, the general Friedmann, acceleration, and continuity equations are given by Eqs. (\ref{eq:General_FRW01_f}), (\ref{eq:General_FRW02_g}), and (\ref{eq:drho_General0}), respectively.
In order to discuss a unified formulation in a matter-dominated universe ($p=0$), we consider the continuity equation written as
\begin{equation}
      \dot{\rho} + 3  \frac{\dot{a}}{a}   \rho   =   U \rho  , 
\label{eq:drho_U_(p=0)}
\end{equation}
where $U$ is given by
\begin{equation} 
 U = 
    \begin{cases} 
         Q            &   (\Lambda(t)    \hspace{1mm} \rm{type}) ,         \\
         \Gamma  &   (\rm{BV} \hspace{1mm} \rm{type}) .                 \\
    \end{cases}
\label{eq:U_Q_Gamma}
\end{equation}
$Q$ [Eq.\ (\ref{eq:Q_f=g})] and $\Gamma$ [Eq.\ (\ref{eq:gamma_CCDM})] are written as
\begin{equation}
  Q =  - \frac{3}{8 \pi G}  \frac{ \dot{f}(t) }{  \rho  }  , 
\label{eq:Q_Unif}
\end{equation}
\begin{equation}
\Gamma = \frac{3}{4 \pi G}  \frac{H g(t)}{ \rho }      .
\label{eq:gamma_Unif_BV}
\end{equation}
Basic hydrodynamical equations for the neo-Newtonian approach are shown in Refs.\ \cite{Lima_Newtonian_1997,Lima2011}.
The basic equations are suitable for describing the BV type.
However, it is necessary to consider the $\Lambda (t)$ type as well, in order to discuss the unified formulation.
Therefore, in the present study, we take into account the fundamental equations for the $\Lambda (t)$ type (examined in the work of Arcuri and Waga \cite{Waga1994}) as well.
Consequently, the basic hydrodynamical equations for the unified formulation can be written as 
\begin{equation}
 \left ( \frac{ \partial \mathbf{u} }{ \partial t } \right )_{r} + ( \mathbf{u} \cdot  \nabla_{r} ) \mathbf{u}   =  -  \nabla_{r} \Phi - \frac{ \nabla_{r} p_{c} } { \rho + \frac{ p_{c} }{ c^{2} } }   , 
\label{eq:Newtonian_1}
\end{equation}
\begin{equation}
 \left ( \frac{ \partial \rho }{ \partial t } \right )_{r} +  \nabla_{r}  \cdot ( \rho \mathbf{u} ) +  \Theta =  0  , 
\label{eq:Newtonian_2_a}
\end{equation}
\begin{equation}
\nabla_{r} ^{2} \Phi = 4 \pi G  \left ( \rho   +  l \right )   , 
\label{eq:Newtonian_3_a}
\end{equation}
where $\mathbf{u}$ is the velocity of a fluid element of volume and $\Phi$ is the gravitational potential.
For the unified formulation, $\Theta$ and $l$ are given as  
\begin{equation} 
 \Theta    =
    \begin{cases} 
         -  Q  \rho        =       \frac{ 3 \dot{f}(t) }{ 8 \pi G }                            &   (\Lambda(t)    \hspace{1mm} \rm{type})  ,         \\
         \frac{ p_{c} }{ c^{2} }  \nabla_{r} \cdot \mathbf{u}                               &   (\rm{BV} \hspace{1mm} \rm{type})          ,          \\
    \end{cases}
\label{eq:u_Unif_Theta}
\end{equation}
\begin{equation} 
 l    =
    \begin{cases} 
        -  \frac{ \Lambda (t) }{ 4 \pi G }       =     -  \frac{ 3 f(t) }{ 4 \pi G }               &   (\Lambda(t)    \hspace{1mm} \rm{type})  ,         \\
        \frac{ 3 p_{c} }{ c^{2} }                                                                       &   (\rm{BV} \hspace{1mm} \rm{type})          .         \\
    \end{cases}
\label{eq:u_Unif_l}
\end{equation}
Equations (\ref{eq:Newtonian_1})--(\ref{eq:Newtonian_3_a}) correspond to the Euler, continuity, and Poisson equations, respectively. 
Using the basic hydrodynamical equations, we have calculated the time evolution equation for the matter density contrast $\delta$, according to the work of Jesus \textit{et al.} \cite{Lima2011}. 
The derivation of the equation is essentially the same as the derivation shown in Ref.\ \cite{Lima2011}. 
(For details, see Ref.\ \cite{Lima2011}.)  
Therefore, we do not discuss this in the present paper. 
Alternatively, we will examine whether the obtained equation is consistent with the equations for the $\Lambda(t)$ and BV types.

Setting $c=1$, using the linear approximation, and neglecting extra terms, the time evolution equation for $\delta$ can be written as
\begin{align}
\ddot{\delta}  & + \left [ H (2  +  3 c_{\rm{eff}}^{2} - 3 u )   - \frac{ \dot{w_{c}} }{ 1+w_{c} }  \right ] \dot{\delta}        \notag \\
                     & +  \Bigg \{      3 (\dot{H} + 2 H^{2}) \left (   c_{\rm{eff}}^{2}  - u \right )     \notag \\
                     & + 3 H  \left [ \dot{c}_{\rm{eff}}^{2}  -  \dot{u}  -  \frac{ \dot{w}_{c} }{ 1+w_{c}  } ( c_{\rm{eff}}^{2} - u )   \right ] \notag \\     
                     &  - 4 \pi G \rho \left ( 1 + w_{c} \right ) ( 1 + 3 c_{\rm{eff}}^{2} )  + \frac{  k^{2} c_{\rm{eff}}^{2} }{ a^{2} }    \Bigg \}     \delta = 0              , 
\label{eq:delta-t_Unif}
\end{align}
where $u$, $w_{c}$, and $c_{\rm{eff}}^{2}$ are defined by 
\begin{equation} 
 u    \equiv  - \frac{U}{3H}      =
    \begin{cases} 
         - \frac{Q}{3H}                                &   (\Lambda(t)    \hspace{1mm} \rm{type})  ,         \\
         - \frac{\Gamma}{3H}  (= w_{c})        &   (\rm{BV} \hspace{1mm} \rm{type})          ,          \\
    \end{cases}
\label{eq:u_Unif_LCDM_BV}
\end{equation}
\begin{equation} 
 w_{c}  \equiv  - \frac{\Gamma}{3H} =
    \begin{cases} 
         0                                       &   (\Lambda(t)    \hspace{1mm} \rm{type})  ,         \\
         - \frac{\Gamma}{3H}           &   (\rm{BV} \hspace{1mm} \rm{type})           ,         \\
    \end{cases}
\label{eq:wc_Unif_LCDM_BV}
\end{equation}
\begin{equation}
 c_{\rm{eff}}^{2}  \equiv   \frac{\delta p_{c} }{\delta \rho }  =
    \begin{cases} 
         0                                                             &   (\Lambda(t)    \hspace{1mm} \rm{type}) ,         \\
         \frac{\delta p_{c} }{\delta \rho }                 &   (\rm{BV} \hspace{1mm} \rm{type}) .                 \\
    \end{cases}
\label{eq:ceff_Unif_LCDM_BV}
\end{equation}
Equation (\ref{eq:delta-t_Unif}) is the unified equation.
In Eq.\ (\ref{eq:delta-t_Unif}), $\rho$ represents $\bar{\rho}$, i.e., a homogenous and isotropic solution for the unperturbed equations. 
For simplicity, we set $c=1$ and replace $\bar{\rho}$ with $\rho$ when we present the time evolution equation for $\delta$.
Note that the values of the three parameters ($u$, $w_{c}$, and $c_{\rm{eff}}^{2}$) for the $\Lambda (t)$ type are different from those for the BV type, as shown in Eqs.\ (\ref{eq:u_Unif_LCDM_BV})--(\ref{eq:ceff_Unif_LCDM_BV}) and discussed in the following paragraph.

In the unified formulation, we consider an effective pressure $p_{e}$ because the BV type assumes the effective pressure, 
where $p_{e}$ is given by $p_{e}   =  p + p_{c} = p_{c} $ in the matter-dominated universe.
For the BV type, we interpret $p_{c}$ as a pressure derived from an entropic-force on the horizon of the universe.
As discussed in Sec.\ \ref{Density perturbations for BV types}, $p_{c}$ for the BV type can be expressed as $ p_{c} = - \rho c^2 \Gamma /(3H)$ [Eq.\ (\ref{eq:pc_CCDM})]. 
Therefore, the equation of state parameter for the BV type is given by $w_{c} = - \Gamma /(3H)$.
In contrast, for the $\Lambda (t)$ type, we neglect the effective pressure, i.e., $p_{c} = 0$.
Therefore, the effective pressure for the two types is written as 
\begin{equation} 
 p_{c} = 
    \begin{cases} 
         0                                                      &   (\Lambda(t)    \hspace{1mm} \rm{type})  ,         \\
         - \frac{\rho c^2 \Gamma}{3H}             &   (\rm{BV} \hspace{1mm} \rm{type})          .         \\
    \end{cases}
\label{eq:pc_Unif_LCDM_BV}
\end{equation}
This indicates that $c_{\rm{eff}}^{2}$, $\Gamma$, and $w_{c}$ are $0$ when we consider the $\Lambda (t)$ type in the matter-dominated universe. 
Consequently, the three parameters are summarized as shown in Eqs.\ (\ref{eq:u_Unif_LCDM_BV})--(\ref{eq:ceff_Unif_LCDM_BV}).
(As discussed in Secs.\ \ref{BV-H model} and \ref{BV-Cst model}, we assume $c_{\rm{eff}}^{2} =0$ for the BV type. 
However, in this subsection, we leave $c_{\rm{eff}}^{2}$ in Eq.\ (\ref{eq:delta-t_Unif}), in order to clarify the difference between the $\Lambda (t)$ and BV types.)

We now examine the unified equation.
To this end, we first consider the $\Lambda (t)$ type.
As discussed above, we can neglect the effective pressure for the $\Lambda (t)$ type. 
Accordingly, $c_{\rm{eff}}^{2}$ and $w_{c}$ are neglected as well.
Substituting $w_{c}=0$, $\dot{w}_{c}=0$, $ c_{\rm{eff}}^{2} =0$, and $\dot{c}_{\rm{eff}}^{2} =0 $ into Eq.\ (\ref{eq:delta-t_Unif})  gives 
\begin{align}
\ddot{\delta}  & + \left [ H (2  - 3 u )   \right ] \dot{\delta}        \notag \\
                     & + \left [   -  3 u (\dot{H} + 2 H^{2})  -  3 H  \dot{u}    - 4 \pi G \rho     \right ]      \delta = 0              ,
\label{eq:delta-t_Unif_LCDM_0}
\end{align}
where $u$ for the $\Lambda (t)$ type [Eq.\ (\ref{eq:u_Unif_LCDM_BV})] is given by
\begin{equation}
 u = - \frac{Q}{3H}  \quad \textrm{and therefore} \quad  \dot{u} = -  \frac{ \dot{Q} H - Q \dot{ H } }{  3 H^{2}  }   .    
\label{eq:u_dotu_Unif_LCDM}
\end{equation}
Substituting Eq.\ (\ref{eq:u_dotu_Unif_LCDM}) into Eq.\ (\ref{eq:delta-t_Unif_LCDM_0}), and rearranging, we have 
\begin{equation}
      \ddot{\delta} +  \left ( 2 H + Q \right ) \dot{\delta} - \left [ 4 \pi G \rho  - 2 H Q -\dot{Q} \right ] \delta = 0           , 
\label{eq:density_Unif_LCDM_1}
\end{equation}
where the mass density $\rho$ for the $\Lambda (t)$ type (in a homogeneous, isotropic, and spatially flat universe) is given by the general Friedmann equation: 
\begin{equation}
 \rho =    \frac{3}{8 \pi G}  (H^{2} -f(t) )           \quad  (\Lambda(t)    \hspace{1mm} \rm{type})     .        
\label{eq:rho_Unif}
\end{equation}
The obtained equation [Eq.\ (\ref{eq:density_Unif_LCDM_1})] is equivalent to Eq.\ (\ref{eq:density_f=g}).
That is, the unified equation recovers the equation for the $\Lambda(t)$ type discussed in Sec.\ \ref{Density perturbations for Lambda(t) types}.
In the following, we use Eq.\ (\ref{eq:density_f=g}) [Eq.\ (\ref{eq:density_Unif_LCDM_1})], in order to examine density perturbations of the $\Lambda(t)$ type.

When we consider the BV type, $u$ is replaced by $w_{c} $ [Eq.\ (\ref{eq:u_Unif_LCDM_BV})].
Substituting $u= w_{c} $ and $\dot{u} = \dot{w}_{c}$ into Eq.\ (\ref{eq:delta-t_Unif}), and rearranging, we have
\begin{align}
\ddot{\delta}  & + \left [ H (2  +  3 c_{\rm{eff}}^{2} - 3 w_{c} )   - \frac{ \dot{w_{c}} }{ 1+w_{c} }  \right ] \dot{\delta}        \notag \\
                     & +  \Bigg \{      3 (\dot{H} + 2 H^{2}) \left (   c_{\rm{eff}}^{2}  - w_{c} \right )     \notag \\
                     & + 3 H  \left [ \dot{c}_{\rm{eff}}^{2}   - ( 1 + c_{\rm{eff}}^{2} )  \frac{ \dot{w}_{c} }{ 1+w_{c}  } \right ] \notag \\     
                     &  - 4 \pi G \rho \left ( 1 + w_{c} \right ) ( 1 + 3 c_{\rm{eff}}^{2} )  + \frac{  k^{2} c_{\rm{eff}}^{2} }{ a^{2} }    \Bigg \}     \delta = 0              , 
\label{eq:delta-t_Unif_BV_0}
\end{align}
where $w_{c}$ for the BV type [Eq.\ (\ref{eq:wc_Unif_LCDM_BV})] is given by
\begin{equation}
 w_{c}= - \frac{\Gamma}{3H}  \quad \textrm{and therefore} \quad \frac{\dot{w}_{c} }{ 1+ w_{c} } =  \frac{ \Gamma \dot{H} - H \dot{\Gamma} }{  H  (3H - \Gamma) }   .    
\label{eq:wc_dotwc_Unif_BV}
\end{equation}
We can confirm that Eq.\ (\ref{eq:delta-t_Unif_BV_0}) is more complicated than Eq.\ (\ref{eq:delta-t_Unif_LCDM_0}), due to extra terms based on an effective pressure.
Substituting Eq.\ (\ref{eq:wc_dotwc_Unif_BV})  into Eq.\ (\ref{eq:delta-t_Unif_BV_0}), we obtain 
\begin{align}
\ddot{\delta}  & + \left [ 2 H + \Gamma + 3 c_{\rm{eff}}^{2} H  - \frac{ \Gamma \dot{H} - H \dot{\Gamma} }{ H (3H -\Gamma)}  \right ] \dot{\delta}        \notag \\
                     & +  \Bigg \{      3 (\dot{H} + 2 H^{2}) \left (   c_{\rm{eff}}^{2}  + \frac{ \Gamma }{ 3H } \right )     \notag \\
                     & + 3 H  \left [ \dot{c}_{\rm{eff}}^{2}   - ( 1 + c_{\rm{eff}}^{2}) \frac{ \Gamma \dot{H} - H \dot{\Gamma} }{ H (3H -\Gamma)}  \right ] \notag \\     
                     &  - 4 \pi G \rho \left ( 1 -  \frac{ \Gamma }{ 3H } \right ) ( 1 + 3 c_{\rm{eff}}^{2} )  + \frac{  k^{2} c_{\rm{eff}}^{2} }{ a^{2} }    \Bigg \}     \delta = 0              , 
\label{eq:delta-t_Unif_BV_1}
\end{align}
where $\rho$ for the BV type (in a homogeneous, isotropic, and spatially flat universe) is given by
\begin{equation}
 \rho =    \frac{3}{8 \pi G}  H^{2}           \quad  (\rm{BV}  \hspace{1mm} \rm{type})     ,        
\label{eq:rho_Unif}
\end{equation}
because $f(t)=0$. 
Equation (\ref{eq:delta-t_Unif_BV_1}) is equivalent to Eq.\ (\ref{eq:delta-t_CCDM}) for the BV type shown in Sec.\ \ref{Density perturbations for BV types}.
Therefore, we can obtain Eqs.\ (\ref{eq:delta-eta_c=0_CCDM_0})--(\ref{eq:G(eta)_0}) from the unified equation.
In the following, we use Eqs.\ (\ref{eq:delta-eta_c=0_CCDM_0})--(\ref{eq:G(eta)_0}), in order to examine density perturbations of the BV type.

In this subsection, we proposed a unified formulation for the $\Lambda (t)$ and BV types, using a neo-Newtonian approach.
The unified formulation considered here can help to discuss density perturbations of the $\Lambda (t)$ and BV types systematically.
Of course, it is possible to examine the density perturbations of the $\Lambda(t)$ and BV types separately, as shown in Secs.\ \ref{Density perturbations for Lambda(t) types} and \ref{Density perturbations for BV types}.

\section{The four entropic-force models}
\label{The present models}

In Sec.\ \ref{Entropic-force models}, we presented the $\Lambda(t)$ and BV types of entropic-force models. 
In Sec.\ \ref{An extended entropic-force model}, we proposed the $H$ and $C_{\rm{cst}}$ versions which include $H$ and constant entropic-force terms, respectively. 
Therefore, in the present paper, we examine four models obtained from combining the $H$ and $C_{\rm{cst}}$ versions with the $\Lambda(t)$ and BV types.
The four models, $\Lambda(t)$-$H$, $\Lambda(t)$-$C_{\rm{cst}}$, BV-$H$, and BV-$C_{\rm{cst}}$, are summarized in Table\ \ref{tab-four models}.

\begin{table}[b]
\caption{The four entropic-force models. The four models are obtained from combining the $H$ and $C_{\rm{cst}}$ versions with the $\Lambda(t)$ and BV types. 
$\hat{\beta_{3}}$ and $\hat{\beta_{4}}$ are dimensional constants defined by Eq. (\ref{eq:ab3ab4-H0_(2)}). }
\label{tab-four models}
\newcommand{\m}{\hphantom{$-$}}
\newcommand{\cc}[1]{\multicolumn{1}{c}{#1}}
\renewcommand{\tabcolsep}{2.5pc} 
\renewcommand{\arraystretch}{1.25} 
\begin{tabular}{@{}lllll}
\hline
\hline
$\textrm{Model}$                          &  $f(t)$                           & $g(t)$      \\
\hline
 $\Lambda(t)$-$H$                       &  $ \hat{\beta}_{3} H$       & $ \hat{\beta}_{3} H$        \\   
 $\Lambda(t)$-$C_{\rm{cst}}$        &  $ \hat{\beta}_{4} $         & $ \hat{\beta}_{4} $          \\   
 BV-$H$                                       &  $0$                              & $ \hat{\beta}_{3} H$        \\   
 BV-$C_{\rm{cst}}$                         &  $0$                             & $ \hat{\beta}_{4} $          \\   
\hline
\hline
\end{tabular}\\
 \end{table}

To examine the four models, we consider a matter-dominated universe given by
\begin{equation}
  w=0 . 
\label{eq:w=0}
\end{equation}
As mentioned in Sec.\ \ref{An extended entropic-force model}, we neglect $H^2$ terms in the entropic-force terms, i.e., $\alpha_1 = \beta_1 = 0$. 
Substituting $\alpha_1 = \beta_1 = 0$ and $w=0$ into Eq.\ (\ref{eq:C1}) gives $C_{1} = 1.5$.
We define $C_{1}$ for a matter-dominated universe as $C_{m}$, which is given by
\begin{equation}
C_{1,m} \equiv C_{m}  = 1.5  \quad  (=3/2) .
\label{eq:Cm=1.5}
\end{equation}
In the standard cosmology, the universe for $C_{1} =3/2$ corresponds to a matter-dominated universe \cite{Weinberg1,Roy1}.

We present formulations of the four models in the following subsections.
(We note that the background evolution of the universe in the four models depends on the two equations Eqs.\ (\ref{eq:dHC1C3hC4h(H)}) and (\ref{eq:dHC1C3hC4h(Cst)}).
Therefore, the background evolution in the $\Lambda(t)$ type is the same as that in the BV type.)

\subsection{$\Lambda(t)$-$H$ model} 
\label{Lambda(t)-H model} 

For the $\Lambda(t)$-$H$ model, the general functions are written as
\begin{equation}
  f(t) = g(t) =  \hat{\beta}_{3}   H           .
\label{eq:fg(H)}
\end{equation}
As shown in Eq.\ (\ref{eq:H/H0_(C1C3)_00}), the evolution of the Hubble parameter for the $H$ version can be rearranged as 
\begin{align}
E (\tilde{a}) \equiv \frac{H} {H_{0}}  &=  \left ( 1-  \frac{ C_{3} }{ C_{m} }  \right )   \left ( \frac{ a } {  a_{0} } \right )^{ -C_{m}}   +  \frac{ C_{3} }{ C_{m} }          \notag \\
                                                    & =  \left ( 1-  \frac{ C_{3} }{ C_{m} }  \right )   \tilde{a}^{ -C_{m}}   +  \frac{ C_{3} }{ C_{m} }          ,
\label{eq:H/H0_(CmC3)_00}
\end{align}
where $E (\tilde{a})$ represents the normalized Hubble parameter $H/H_{0}$ and
$\tilde{a}$ is the normalized scale factor $a/a_{0}$.
In Eq.\ (\ref{eq:H/H0_(CmC3)_00}),  $C_{1}$ has been replaced by $C_{m} (=3/2)$ [Eq.\ (\ref{eq:Cm=1.5})].

In the present study, $C_{3}$ is determined through fitting with a fine-tuned standard $\Lambda$CDM model, as discussed in Sec. \ref{Results}. 
Using the obtained $C_{3}$ and substituting $w=0$ and $\alpha_{3} = \beta_{3}$ into Eq.\ (\ref{eq:C3}), we have 
\begin{equation}
\beta_{3} = \frac{2}{3}  C_{3}  .   
\label{eq:a3b3_L-H_model}
\end{equation}
Similarly, $\hat{\beta}_3$ is given by $(2/3) C_{3} H_{0}$.

We now discuss density perturbations.
To this end, we employ density perturbations of the $\Lambda(t)$ type shown in Sec.\ \ref{Density perturbations for Lambda(t) types}.
(For a unified formulation based on a neo-Newtonian approach, see Sec.\ \ref{Unified formulation}.)  
Substituting Eq.\ (\ref{eq:fg(H)}) into Eq.\ (\ref{eq:drho_General0_f=g_(p=0)}), we obtain \cite{C11}
\begin{equation}
      \dot{\rho} + 3  \frac{\dot{a}}{a}   \rho   =  - \frac{3}{8 \pi G} \hat{\beta}_{3} \dot{H}   .
\label{eq:drho_General0_f=g_(p=0)_L-H}
\end{equation}
Consequently, the time evolution equation for the perturbation growth factor in the $\Lambda(t)$-$H$ model is given by 
\begin{equation}
      \ddot{\delta} +  \left ( 2 H + Q \right ) \dot{\delta} - \left [ 4 \pi G \rho  - 2 H Q  -\dot{Q}  \right ] \delta = 0 ,
\label{eq:density_f=g_L-H}
\end{equation}
\begin{equation}
 \rho =   \frac{3}{8 \pi G}  (H^{2} - \hat{\beta}_{3} H ),     \quad   Q  =  - \frac{3}{8 \pi G}  \frac{ \hat{\beta}_{3} \dot{H} }{  \rho  }  .
\label{eq:Q_f=g_L-H}
\end{equation}
The above equations are equivalent to a `$\Lambda \propto H $ model' examined by Basilakos \textit{et al.} \cite{Sola_2009}.
Based on their solutions,  the perturbation growth factor is written as
\begin{equation} 
   \delta (\tilde{a}) =   J  \tilde{a}^{-3/2} \int^{\tilde{a}}_{0}   \frac{ dx }{ x^{3/2}  E(x)^{2} }   ,
\label{eq:delta_L-H}
\end{equation}
where  $J$ is given by 
\begin{align}
J  &= \frac{3}{2}  \left ( 1- \frac{C_{3}}{C_{m}} \right )^{2}  \left ( \frac{  \frac{C_{3}}{C_{m}} }{ 1- \frac{C_{3}}{C_{m}} } \right )^{2/3}   \notag \\
   &= \frac{3}{2}  \left ( 1- \frac{C_{3}}{C_{m}} \right )^{4/3}  \left (  \frac{C_{3}}{C_{m}}  \right )^{2/3}    .
\label{eq:J_L-H}
\end{align}
$E(x)$ for the $H$ version is given by Eq.\ (\ref{eq:H/H0_(CmC3)_00}).
Therefore, we can examine the perturbation growth factor for the $\Lambda(t)$-$H$ model using the determined $C_{3}$ and $C_{m} =3/2$.
We note that $C_{3} /C_{m}$ included in Eqs.\ (\ref{eq:H/H0_(CmC3)_00}) and (\ref{eq:J_L-H}) behaves as if it were the density parameter $\Omega_{\Lambda}$ for $\Lambda$ in the $\Lambda \propto H $ model \cite{Sola_2009}.  
Similarly, $1- \frac{C_{3}}{C_{m}}$ behaves as if it were the density parameter $\Omega_{m}$ for matter in the $\Lambda \propto H $ model.
Note that $\Omega_{m}$ is given by $1- \Omega_{\Lambda} $, when we assume a flat universe and neglect the density parameter $\Omega_{r}$ for the radiation. 
In this study, the $C_{3} /C_{m}$ term depends on an entropic-force derived from a volume entropy.

\subsection{$\Lambda(t)$-$C_{\rm{cst}}$ model} 
\label{Lambda(t)-Cst model} 

For the $\Lambda(t)$-$C_{\rm{cst}}$ model, the general functions are written as
\begin{equation}
  f(t) =   g(t) =    \hat{\beta}_{4}     .
\label{eq:fg(Cst)}
\end{equation}
The formulation of this model is equivalent to the standard $\Lambda$CDM model. 
As shown in Eq.\ (\ref{eq:H/H0(C1C4)_00}), the evolution of the Hubble parameter for the $C_{\rm{cst}}$ version is given as 
\begin{equation}
E (\tilde{a}) ^{2} = \left ( \frac{H} {H_{0}} \right )^{2}   
=  \left ( 1-  \frac{ C_{4} }{ C_{m} }  \right )  \tilde{a}^{ -2 C_{m}}   +  \frac{ C_{4} }{ C_{m} }     , 
\label{eq:H/H0(CmC4)_00}
\end{equation}
where $C_{1}$ has been replaced by $C_{m} (=3/2)$ and $\tilde{a}$ represents $a/a_{0}$.
This equation is equivalent to the solution of the standard $\Lambda$CDM model.
Therefore, the constant term $C_{4}/C_{m}$ behaves as if it were $\Omega_{\Lambda}$ in the standard $\Lambda$CDM model.
Similarly, $1- \frac{C_{4}}{C_{m}}$ behaves as if it were $\Omega_{m}$ in the standard $\Lambda$CDM model in a flat universe. 
In this study, we determine $C_{4}$ from $\Omega_{\Lambda}$ of a fine-tuned standard $\Lambda$CDM model, as discussed in Sec.\ \ref{Results}. 
Accordingly, we can obtain $\beta_{4}$ from the determined $C_{4}$. 
Substituting $w=0$ and $\alpha_{4} = \beta_{4}$ into Eq.\ (\ref{eq:C4}), $\beta_{4}$ is written as
\begin{equation}
\beta_{4} = \frac{2}{3}  C_{4}  .  
\label{eq:a4b4_L-Cst_model}
\end{equation}
Similarly, $\hat{\beta}_4$ is given by $(2/3) C_{4} H_{0}^{2}$.

In order to discuss density perturbations of the $\Lambda(t)$ type, we substitute Eq.\ (\ref{eq:fg(Cst)}) into Eqs.\ (\ref{eq:drho_General0_f=g_(p=0)}) and (\ref{eq:Q_f=g}) to give
\begin{equation}
      \dot{\rho} + 3  \frac{\dot{a}}{a}   \rho   =  0 ,   \quad Q=0           .
\label{eq:drho_General0_f=g_(p=0)_L-C_Q=0}
\end{equation}
Substituting $Q=0$ into Eq.\ (\ref{eq:density_f=g}), the time evolution equation for the perturbation growth factor is written as
\begin{equation}
      \ddot{\delta} +    2 H  \dot{\delta} -  4 \pi G \rho \delta = 0     .
\label{eq:density_f=g_L-C_model}
\end{equation}
Solving Eq.\ (\ref{eq:density_f=g_L-C_model}), we obtain the well-known perturbation growth factor \cite{Peebles_1993,Sola_2009} given as
\begin{equation} 
   \delta (\tilde{a}) = \frac{  5 \left ( 1- \frac{C_{4}}{C_{m}} \right ) E  (\tilde{a})  }{  2  } \int^{\tilde{a}}_{0}   \frac{ dx }{ x^{3}  E (x)^{3} }    ,
\label{eq:delta_L-C_model}
\end{equation}
where $E(x)$ for the $C_{\rm{cst}}$ version is calculated from Eq.\ (\ref{eq:H/H0(CmC4)_00}).
This solution is the same as the standard $\Lambda$CDM model.  
Note that we assume an entropy $S_{r4}$ proportional to $r_{H}^{4}$ in a matter-dominated universe  \cite{C12}.

\subsection{BV-$H$ model} 
\label{BV-H model} 

For the BV-$H$ model, the general functions are written as 
\begin{equation}
  f(t) = 0,      \quad  g(t) =  \hat{\beta}_{3}   H           .
\label{eq:f=0_g=H}
\end{equation}
As shown in Eq.\ (\ref{eq:H/H0_(CmC3)_00}), the evolution of the Hubble parameter for the $H$ version is given by 
\begin{equation}
E (\tilde{a}) \equiv \frac{H} {H_{0}}    =  \left ( 1-  \frac{ C_{3} }{ C_{m} }  \right )   \tilde{a}^{ -C_{m}}   +  \frac{ C_{3} }{ C_{m} }          .
\label{eq:H/H0_(CmC3)_BV-H}
\end{equation}
In the present study, $C_{3}$ is determined through fitting with a fine-tuned standard $\Lambda$CDM model.
Consequently, the obtained $C_{3}$ for the BV-$H$ model is the same as that for the $\Lambda(t)$-$H$ model. 
Substituting $w=0$ and $\alpha_{3} = 0$ into Eq.\ (\ref{eq:C3}), $\beta_{3}$ is given by
\begin{equation}
\beta_{3} =   C_{3}  .   
\label{eq:a3b3_BV-H_model}
\end{equation}
The obtained $\beta_{3}= C_{3}$ is slightly different from $\beta_{3} = (2/3) C_{3}$ [Eq.\ (\ref{eq:a3b3_L-H_model})] for the $\Lambda(t)$-$H$ model.

We now examine the density perturbations.
As discussed in Sec.\ \ref{Density perturbations for BV types}, when $p=0$, the general continuity equation for the BV type [Eq.\ (\ref{eq:drho_General0_f=0_(p=0)})] is written as  
\begin{equation}
      \dot{\rho} + 3  \frac{\dot{a}}{a}   \rho   =   \frac{3}{4 \pi G}  H g(t)   .
\label{eq:drho_General0_f=0_(p=0)01}
\end{equation}
In addition, the parameter $\Gamma$ for entropy production processes [Eq.\ (\ref{eq:gamma_CCDM})] is written as 
\begin{equation}
\Gamma = \frac{3}{4 \pi G}  \frac{H g(t)}{ \rho }      . 
\label{eq:gamma_CCDM_01}
\end{equation}
In the following, we apply the method proposed in Ref.\ \cite{Lima2011} to the BV-$H$ model.

For the BV-$H$ model, $g(t)$ is given by $\hat{\beta}_{3} H (= \beta_{3} H_{0} H)$. 
Substituting Eq.\ (\ref{eq:a3b3_BV-H_model}) into this equation, we have 
\begin{equation}
g(t) = C_{3} H_{0} H       .
\label{eq:g(t)_BV-Cst_C3}
\end{equation}
Substituting Eq.\ (\ref{eq:g(t)_BV-Cst_C3}) into Eq.\ (\ref{eq:gamma_CCDM_01}), $\Gamma$ is given by 
\begin{equation}
\Gamma = \frac{3}{4 \pi G}  \frac{C_{3} H_{0} H^{2}  }{ \rho }      .
\label{eq:gamma_CCDM_01_H}
\end{equation}
Using the critical density $\rho_{c0} = 3 H_{0}^{2}/(8 \pi G)$ and $C_{m}= 3/2$ [Eq.\ (\ref{eq:Cm=1.5})], 
we can rearrange Eq.\ (\ref{eq:gamma_CCDM_01_H}) as
\begin{align}
\Gamma & = \frac{3}{4 \pi G}  \frac{C_{3} H_{0} H^{2}  }{ \rho }    =  2  C_{3} \left (   \frac{ \rho_{c0}  }{ \rho }  \right ) \frac{ H^{2} }{ H_{0} }         \notag \\
             & =  3 \left ( \frac{ C_{3} }{ C_{m} } \right ) \left (   \frac{ \rho_{c0}  }{ \rho }  \right ) \frac{ H^{2} }{ H_{0}}  
                =  3 \tilde{\Omega}_{\Lambda_{H}} \left (   \frac{ \rho_{c0}  }{ \rho }  \right ) \frac{ H^{2} }{ H_{0} }              ,
\label{eq:gamma_CCDM_01_H_Omega}
\end{align}
where $\tilde{\Omega}_{\Lambda_{H}}$ for the BV-$H$ model is given as
\begin{equation}
      \tilde{\Omega}_{\Lambda_{H}}  =  \frac{ C_{3} }{ C_{m} }      .
\label{eq:Omega3}
\end{equation}
We emphasize that $\tilde{\Omega}_{\Lambda_{H}}$ is \textit{not} the density parameter for $\Lambda$, but is a constant parameter, 
although $\tilde{\Omega}_{\Lambda_{H}}$ behaves as if it were the density parameter for $\Lambda$.

When we examine the BV type, the time evolution equation for the perturbation growth factor is given by Eq.\ (\ref{eq:delta-t_CCDM}) which includes $c_{\rm{eff}}^{2}$ terms, as shown in Secs.\ \ref{Density perturbations for BV types} and \ref{Unified formulation}. 
In the present study, we assume $c_{\rm{eff}}^{2} = 0$, in order to ensure an equivalence between the neo-Newtonian and general relativistic approaches.
This is because the neo-Newtonian equation [Eq.\ (\ref{eq:delta-t_CCDM})] is equivalent to the general relativistic equation for a single-fluid-dominated universe only when $c_{\rm{eff}}^{2} = 0$, as examined in the work of Reis \cite{Reis_2003}. 
The equivalence is closely discussed in the recent work of Ramos \textit{et al.} \cite{Ramos_2014}.
(Substituting Eq.\ (\ref{eq:gamma_CCDM_01_H_Omega}) into Eq.\ (\ref{eq:pc_CCDM}), we have a time-dependent effective pressure ${p}_{c}$.
Accordingly, $c_{s}^{2}$ is not $0$, where $c_{s}^{2}$ is defined by $c_{s}^{2} \equiv \dot{p}_{c}/\dot{\rho}$.
This implies non-adiabatic perturbations because $c_{\rm{eff}}^{2} \neq c_{s}^{2}$ \cite{Reis_2003,Ramos_2014}.)

As discussed above, in the present study, we consider $c_{\rm{eff}}^{2}  = 0$.
Therefore, Eqs.\ (\ref{eq:delta-eta_c=0_CCDM_0})--(\ref{eq:G(eta)_0}) are written as 
\begin{equation}
\delta^{\prime \prime}  + F(\eta) \delta^{\prime}  +  G(\eta) \delta =0, 
\label{eq:delta-eta_c=0_BV-H}
\end{equation}
where 
\begin{equation}
F(\eta) =  2  + \frac{ \Gamma + H^{\prime} }{ H }   -   \frac{  \Gamma H^{\prime} - H \Gamma^{\prime}  }{ H (3H -\Gamma) }   ,
\label{eq:F(eta)_c=0_BV-H}
\end{equation}
\begin{equation}
G(\eta)  =  \left (  \frac{ \Gamma }{ H }  - 1   \right )   \left (  \frac{ \Gamma }{ 2H }  + \frac{3}{2}   \right )   -  3 \frac{  \Gamma H^{\prime} - H \Gamma^{\prime}  }{ H (3H -\Gamma) }   .
\label{eq:G(eta)_c=0_BV-H}
\end{equation}
In Eqs.\ (\ref{eq:delta-eta_c=0_BV-H})--(\ref{eq:G(eta)_c=0_BV-H}), $^{\prime}$ represents the differential with respect to $\eta$, i.e., $d/d \eta$, where $\eta \equiv \ln (\tilde{a}(t))$  [Eq.\ (\ref{eta_def})].
It should be noted that $H$ and $\Gamma$ included in Eqs.\ (\ref{eq:delta-eta_c=0_BV-H})--(\ref{eq:G(eta)_c=0_BV-H}) for the BV-$H$ model are different from those for the LJO model \cite{Lima2011}.

From Eqs.\ (\ref{eq:H/H0_(CmC3)_BV-H}), (\ref{eq:gamma_CCDM_01_H_Omega}), and (\ref{eq:Omega3}),   $\Gamma / H$ can be rearranged as 
\begin{align}
\frac{\Gamma}{H} &=   3 \tilde{\Omega}_{\Lambda_{H}} \left (   \frac{ \rho_{c0}  }{ \rho }  \right ) \frac{ H }{ H_{0} }  
                             =   3 \tilde{\Omega}_{\Lambda_{H}} \left (   \frac{ H_{0} }{H }  \right )^{2} \frac{ H }{ H_{0} }                 \notag \\
                           &= \frac{ 3 \tilde{\Omega}_{\Lambda_{H}}  }{  ( 1-  \tilde{\Omega}_{\Lambda_{H}}  )   \tilde{a}^{ -C_{m}}   +  \tilde{\Omega}_{\Lambda_{H}}                }
                             = \frac{ 3 \tilde{\Omega}_{\Lambda_{H}} \tilde{a}^{C_{m}}  }{  ( 1-  \tilde{\Omega}_{\Lambda_{H}}  )     +  \tilde{\Omega}_{\Lambda_{H}}   \tilde{a}^{C_{m}}              }         \notag \\                
                           &= \frac{ 3 \tilde{\Omega}_{\Lambda_{H}}  e^{\frac{3}{2}\eta}  }{   1-  \tilde{\Omega}_{\Lambda_{H}}      +  \tilde{\Omega}_{\Lambda_{H}}   e^{\frac{3}{2}\eta}              }    ,
\label{eq:Gamma-H_c0_BV-H}
\end{align}
where $\rho_{c0} / \rho$ is replaced by $(H_{0}/H)^{2}$ using the Friedmann equation.
Also, $\tilde{a}^{C_{m}}$ is replaced by $e^{3 \eta /2}$ using $C_{m} =3/2$ and $ \tilde{a} = e^{\eta}$.
Similarly, we obtain 
\begin{equation}
\frac{\Gamma + H^{\prime} }{H} = \frac{ -\frac{3}{2} ( 1- \tilde{\Omega}_{\Lambda_{H}} )   + 3 \tilde{\Omega}_{\Lambda_{H}}  e^{\frac{3}{2}\eta}  }{   1-  \tilde{\Omega}_{\Lambda_{H}}      +  \tilde{\Omega}_{\Lambda_{H}}   e^{\frac{3}{2}\eta}              }    ,
\label{eq:Gamma-H_c0_BV-H_2}
\end{equation}
\begin{equation}
 \frac{  \Gamma H^{\prime} - H \Gamma^{\prime}  }{ H (3H -\Gamma) }  
      =    \frac{ -\frac{3}{2}  \tilde{\Omega}_{\Lambda_{H}}  e^{\frac{3}{2}\eta}  }{   1-  \tilde{\Omega}_{\Lambda_{H}}  +  \tilde{\Omega}_{\Lambda_{H}}   e^{\frac{3}{2}\eta}              }    . 
\label{eq:Gamma-H_c0_BV-H_3}
\end{equation}
From Eqs. (\ref{eq:Gamma-H_c0_BV-H})--(\ref{eq:Gamma-H_c0_BV-H_3}),  $F(\eta)$ can be arranged  as 
\begin{equation}
F(\eta)=     \frac{  \left ( 1 - \tilde{\Omega}_{\Lambda_{H}} \right )    + 13 \tilde{\Omega}_{\Lambda_{H}}  e^{\frac{3}{2}\eta}                         }
                         { 2 \left ( 1-  \tilde{\Omega}_{\Lambda_{H}}  +  \tilde{\Omega}_{\Lambda_{H}}   e^{\frac{3}{2}\eta}  \right )      }        ,     
\label{eq:F_BV-H_c0_1}
\end{equation}
or
\begin{equation}
F(\eta)=     \frac{    \left ( 1 - \frac{C_{3}}{C_{m}} \right )  +  13 \frac{C_{3}}{C_{m}}  e^{\frac{3}{2}\eta}                           }
                         { 2  \left ( 1-  \frac{C_{3}}{C_{m}}      +  \frac{C_{3}}{C_{m}}   e^{\frac{3}{2}\eta}   \right )  }            ,
\label{eq:F_BV-H_c0_2}
\end{equation}
and $G(\eta)$ can be arranged as 
\begin{align}
G(\eta) =& \frac{    3  \left  \{   4 \tilde{\Omega}_{\Lambda_{H}}^{2}  e^{ 3 \eta}    - \left ( 1 - \tilde{\Omega}_{\Lambda_{H}} \right )^{2}   \right    \}                    }                      
                        {    2 \left ( 1-  \tilde{\Omega}_{\Lambda_{H}}      +  \tilde{\Omega}_{\Lambda_{H}}   e^{\frac{3}{2}\eta}   \right )^{2}         }           \notag \\
             &+ \frac{   9 \tilde{\Omega}_{\Lambda_{H}}  e^{\frac{3}{2}\eta}                                                      }      
                          {   2 \left ( 1-  \tilde{\Omega}_{\Lambda_{H}}      +  \tilde{\Omega}_{\Lambda_{H}}   e^{\frac{3}{2}\eta}    \right )         }     ,
\label{eq:G_BV-H_c0_1}
\end{align}
or 
\begin{align}
G(\eta) =& \frac{    3  \left  \{   4  \left ( \frac{C_{3}}{C_{m}} \right  )^{2}  e^{ 3 \eta}    - \left ( 1 - \frac{C_{3}}{C_{m}}  \right )^{2}   \right    \}                  }                      
                        {    2 \left ( 1-  \frac{C_{3}}{C_{m}}  +  \frac{C_{3}}{C_{m}}  e^{\frac{3}{2}\eta}   \right )^{2}                                                                        }           \notag \\
             &+ \frac{   9 \frac{C_{3}}{C_{m}}    e^{\frac{3}{2}\eta}                                                      }      
                          {   2 \left ( 1-  \frac{C_{3}}{C_{m}}  +  \frac{C_{3}}{C_{m}}  e^{\frac{3}{2}\eta}    \right )         }     .
\label{eq:G_BV-H_c0_2}
\end{align}
Using $F(\eta)$ and $G(\eta)$, we can numerically solve the differential equation [Eq.\ (\ref{eq:delta-eta_c=0_BV-H})] for the BV-$H$ model. 
To solve this, we employ the initial conditions of the Einstein--de Sitter growing model \cite{Lima2011}. 
The initial conditions are set to be $\delta (\tilde{a}_{i}) = \tilde{a}_{i}$ and $\delta^{\prime}  (\tilde{a}_{i}) = \tilde{a}_{i}$, where $\tilde{a}_{i} =a_{i}/a_{0} = 10^{-3}$.

\subsection{BV-$C_{\rm{cst}}$ model} 
\label{BV-Cst model} 

For the BV-$C_{\rm{cst}}$ model, the general functions are written as
\begin{equation}
  f(t) = 0,      \quad g(t) =  \hat{\beta}_{4}            .
\label{eq:f=0_g=Cst}
\end{equation}
As shown in Eq.\ (\ref{eq:H/H0(CmC4)_00}),  the evolution of the Hubble parameter for the $C_{\rm{cst}}$ version is given by 
\begin{equation}
E (\tilde{a}) ^{2} = \left ( \frac{H} {H_{0}} \right )^{2}   
=  \left ( 1-  \frac{ C_{4} }{ C_{m} }  \right )  \tilde{a}^{ -2 C_{m}}   +  \frac{ C_{4} }{ C_{m} }     .
\label{eq:H/H0(CmC4)_BV-Cst}
\end{equation}
In the present study, we determine $C_{4}$ from $\Omega_{\Lambda}$ of a fine-tuned standard $\Lambda$CDM model, as discussed in Sec.\ \ref{Results}.
That is, the obtained $C_{4}$ for the BV-$C_{\rm{cst}}$ model is the same as $C_{4}$ for the $\Lambda(t)$-$C_{\rm{cst}}$ model. 
Substituting $w=0$ and $\alpha_{4} = 0$ into Eq.\ (\ref{eq:C4}), $\beta_{4}$ is given by
\begin{equation}
\beta_{4} =   C_{4}  .   
\label{eq:a4b4_BV-Cst_model}
\end{equation}
The obtained $\beta_{4}= C_{4}$ is slightly different from $\beta_{4} = (2/3) C_{4}$ [Eq.\ (\ref{eq:a4b4_L-Cst_model})] for the $\Lambda(t)$-$C_{\rm{cst}}$ model.

We now examine the density perturbations.
When $p=0$, the general continuity equation for the BV type [Eq.\ (\ref{eq:drho_General0_f=0_(p=0)})] is written as 
\begin{equation}
      \dot{\rho} + 3  \frac{\dot{a}}{a}   \rho   =   \frac{3}{4 \pi G}  H g(t)   , 
\label{eq:drho_General0_f=0_(p=0)2}
\end{equation}
and the parameter $\Gamma$ [Eq.\ (\ref{eq:gamma_CCDM})] is written as  
\begin{equation}
\Gamma = \frac{3}{4 \pi G}  \frac{H g(t)}{ \rho }      .
\label{eq:gamma_CCDM_02}
\end{equation}
For the BV-$C_{\rm{cst}}$ model, $g(t)$ is given by $\hat{\beta}_{4}= \hat{C}_{4}= C_{4} H_{0}^{2}$. 
Substituting $g(t) = C_{4} H_{0}^{2}$, $\rho_{c0} = 3 H_{0}^{2}/(8 \pi G)$, and $C_{m}= 3/2$ into Eq.\ (\ref{eq:gamma_CCDM_02}) and rearranging, we have 
\begin{align}
\Gamma & = \frac{3}{4 \pi G}  \frac{ H (C_{4} H_{0}^{2})  }{ \rho }      =  2  C_{4} \left (   \frac{ \rho_{c0}  }{ \rho }  \right )  H        \notag \\
             & =  3 \left ( \frac{ C_{4} }{ C_{m} } \right ) \left (   \frac{ \rho_{c0}  }{ \rho }  \right ) H
                =  3 \tilde{\Omega}_{\Lambda}                \left (   \frac{ \rho_{c0}  }{ \rho }  \right ) H              ,
\label{eq:gamma_CCDM_01_Cst_Omega}
\end{align}
where $\tilde{\Omega}_{\Lambda}$ for the BV-$C_{\rm{cst}}$ model is given as
\begin{equation}
\tilde{\Omega}_{\Lambda}  =  \frac{C_{4}}{C_{m}}     .
\label{eq:pc_cst_CCDM_Omega-L}
\end{equation}
Note that $\tilde{\Omega}_{\Lambda}$ is \textit{not} the density parameter for $\Lambda$, but is a constant parameter, 
although $\tilde{\Omega}_{\Lambda}$ behaves as if it were the density parameter for $\Lambda$.
The above equations are equivalent to those for the LJO model \cite{Lima2010,Lima2011}.
For example, Eq.\ (\ref{eq:gamma_CCDM_01_Cst_Omega}) is the same as Eq.\ (\ref{eq:gamma-H_cst_CCDM}).

When we examine the BV type, the time evolution equation for the perturbation growth factor is given by Eq.\ (\ref{eq:delta-t_CCDM}).
In the present study, we consider $c_{\rm{eff}}^{2}  = 0$, in order to ensure an equivalence between the neo-Newtonian and general relativistic approaches \cite{Reis_2003}, as discussed in Sec.\ \ref{BV-H model}. 
Of course, we can expect $c_{\rm{eff}}^{2}=0$ for the BV-$C_{\rm{cst}}$ model \cite{Lima2011}.
For example, substituting Eq.\ (\ref{eq:gamma_CCDM_01_Cst_Omega}) into Eq.\ (\ref{eq:pc_CCDM}), we obtain a constant effective pressure.
Therefore, we expect that the pressure perturbation $\delta p_{c}$ should vanish, i.e., $c_{\rm{eff}}^{2} \equiv \delta p_{c} /\delta \rho  =0$.
However, strictly speaking, the constant effective pressure is not equivalent to $\delta p_{c} = 0$. 
Accordingly, in the present paper, we explicitly assume $c_{\rm{eff}}^{2} =0$.
(We obtain $c_{s}^{2} = 0$ from the constant effective pressure since $c_{s}^{2}$ is defined by $\dot{p}_{c}/\dot{\rho}$. 
This indicates adiabatic perturbations because $c_{\rm{eff}}^{2} = c_{s}^{2}$ \cite{Reis_2003,Ramos_2014}.
The influence of $c_{\rm{eff}}^{2}$ is closely examined in Ref.\ \cite{Lima2011}.)
Consequently, Eqs.\ (\ref{eq:delta-eta_c=0_CCDM_0})--(\ref{eq:G(eta)_0}) can be written as 
\begin{equation}
\delta^{\prime \prime}  + F(\eta) \delta^{\prime}  +  G(\eta) \delta =0, 
\label{eq:delta-eta_c=0_CCDM}
\end{equation}
where 
\begin{equation}
F(\eta) =  2  + \frac{ \Gamma + H^{\prime} }{ H }   -   \frac{  \Gamma H^{\prime} - H \Gamma^{\prime}  }{ H (3H -\Gamma) }   ,
\label{eq:F(eta)_c=0}
\end{equation}
\begin{equation}
G(\eta)  =  \left (  \frac{ \Gamma }{ H }  - 1   \right )   \left (  \frac{ \Gamma }{ 2H }  + \frac{3}{2}   \right )   -  3 \frac{  \Gamma H^{\prime} - H \Gamma^{\prime}  }{ H (3H -\Gamma) }   .
\label{eq:G(eta)_c=0}
\end{equation}
Substituting Eqs.\ (\ref{eq:H/H0(CmC4)_BV-Cst}), (\ref{eq:gamma_CCDM_01_Cst_Omega}), and (\ref{eq:pc_cst_CCDM_Omega-L}) into Eqs.\ (\ref{eq:F(eta)_c=0}) and (\ref{eq:G(eta)_c=0}) and rearranging, $F(\eta)$ and $G(\eta)$ \cite{Lima2011} can be summarized as%
\begin{equation}
F(\eta) = \frac{  \left ( 1 - \tilde{\Omega}_{\Lambda} \right )  + 16 \tilde{\Omega}_{\Lambda}  e^{3 \eta}           }{  2 \left ( 1 - \tilde{\Omega}_{\Lambda}  +  \tilde{\Omega}_{\Lambda}  e^{3 \eta}  \right )  }  
             = \frac{  \left ( 1 - \frac{C_{4}}{C_{m}}  \right )  + 16 \frac{C_{4}}{C_{m}}   e^{3 \eta}           }{  2  \left ( 1 - \frac{C_{4}}{C_{m}}   +  \frac{C_{4}}{C_{m}}  e^{3 \eta}  \right  )  }                 ,
\label{eq:F_BV_cst_c0}
\end{equation}
\begin{align}
G(\eta) =&  \frac{   3 \left \{ 4 \tilde{\Omega}_{\Lambda}^{2} e^{6 \eta} -   \left ( 1 - \tilde{\Omega}_{\Lambda} \right )^{2} \right \}  }{  2  \left ( 1 - \tilde{\Omega}_{\Lambda}  +  \tilde{\Omega}_{\Lambda}  e^{3 \eta}  \right  )^{2}  }     
                 + \frac{ 9 \tilde{\Omega}_{\Lambda}  e^{3 \eta}    }{    1 - \tilde{\Omega}_{\Lambda}  +  \tilde{\Omega}_{\Lambda}  e^{3 \eta}      }   \notag \\   
           =&  \frac{   9 \left ( 1 - \tilde{\Omega}_{\Lambda} \right )^{2}  }{  2  \left ( 1 - \tilde{\Omega}_{\Lambda}  +  \tilde{\Omega}_{\Lambda}  e^{3 \eta}  \right  )^{2}  }  
                +   \frac{   3 \left ( 5 \tilde{\Omega}_{\Lambda}  e^{3 \eta}  - 2  + 2 \tilde{\Omega}_{\Lambda} \right  )  }{    1 - \tilde{\Omega}_{\Lambda}  +  \tilde{\Omega}_{\Lambda}  e^{3 \eta}      }   \notag \\    
           =&  \frac{   9 \left ( 1 - \frac{C_{4}}{C_{m}}  \right  )^{2}  }{  2 \left ( 1 - \frac{C_{4}}{C_{m}} +  \frac{C_{4}}{C_{m}}  e^{3 \eta}  \right )^{2}  }  
                +   \frac{   3 \left ( 5 \frac{C_{4}}{C_{m}}  e^{3 \eta}  - 2  + 2 \frac{C_{4}}{C_{m}} \right  )        }{    1 - \frac{C_{4}}{C_{m}}  +  \frac{C_{4}}{C_{m}}  e^{3 \eta}      }            .  \notag \\   
\label{eq:G_BV_cst_C0}
\end{align}
The evolution of $\delta$ is solved numerically, using the initial conditions of the Einstein--de Sitter growing model \cite{Lima2011}, as described in Sec.\ \ref{BV-H model}.

\section{Evolution of the universe in the four entropic-force models}
\label{Results}

\begin{table*}[t]
\caption{Dimensionless constants for the four entropic-force models. 
We consider a matter-dominated universe, i.e., $C_{1} = C_{m} =1.5$.
For the $\Lambda(t)$-$H$ and BV-$H$ models, $C_{3}$ is determined through fitting with the luminosity distance of a fine-tuned standard $\Lambda$CDM model \cite{Koma5}.
For the $\Lambda(t)$-$C_{\rm{cst}}$ and BV-$C_{\rm{cst}}$ models, $C_{4}$ is calculated from $C_{1} \Omega_{\Lambda}$, where $\Omega_{\Lambda}$ is $0.685$ based on the Planck 2013 results \cite{Planck2013}. 
For details, see the text. }
\label{tab-parameter}
\newcommand{\m}{\hphantom{$-$}}
\newcommand{\cc}[1]{\multicolumn{1}{c}{#1}}
\renewcommand{\tabcolsep}{2.0pc} 
\renewcommand{\arraystretch}{1.25} 
\begin{tabular}{@{}lllll}
\hline
\hline
$\textrm{Parameter}$   &  $\Lambda(t)$-$H$     & $\Lambda(t)$-$C_{\rm{cst}}$   & BV-$H$              & BV-$C_{\rm{cst}}$        \\
\hline
$\alpha_{3}$                 &  $(2/3) C_{3}$             & $0$                                        &  $0$                  & $0$                          \\
$\alpha_{4}$                 &  $0$                           & $(2/3) C_{4}$                          &  $0$                  & $0$                          \\
$\beta_{3}$                   &  $(2/3) C_{3}$             & $0$                                        &  $C_{3}$             & $0$                          \\
$\beta_{4}$                   &  $0$                           & $(2/3) C_{4}$                          &  $0$                  & $C_{4}$                     \\
\hline
$C_{1}$                         &  $1.5$                         & $1.5$                                     & $1.5$                 & $1.5$                        \\     
$C_{3}$                         &  $0.884$                      & $0$                                       & $0.884$             &  $0$                         \\    
$C_{4}$                         &  $0$                            & $1.03$                                   &  $0$                  &  $1.03$                     \\     
\hline
\hline
\end{tabular}\\
 \end{table*}

We examine the evolution of the universe in the four entropic-force models, $\Lambda(t)$-$H$, $\Lambda(t)$-$C_{\rm{cst}}$, BV-$H$, and BV-$C_{\rm{cst}}$. 
To discuss the properties of the four models, we determine four dimensionless constants, $\alpha_3$, $\alpha_4$, $\beta_3$, and $\beta_4$, from the background evolution of the universe.
The obtained constants for the four models are summarized in Table\ \ref{tab-parameter}.
(Several parameters are $0$ based on our definition of each model.
For example, for the $\Lambda(t)$-$H$ model, both $\alpha_4$ and $\beta_4$ are $0$, because constant terms with $\alpha_4$ and $\beta_4$ are neglected in this model.)

For the $H$ version (i.e., the $\Lambda(t)$-$H$ and BV-$H$ models), $C_{4}$ is $0$. 
Accordingly, for the $H$ version, we determine $C_{3}$ through fitting with a fine-tuned standard $\Lambda$CDM model \cite{Planck2013}.
To this end, we use the luminosity distance, as examined in Ref.\ \cite{Koma5}.
After $C_{3}$ and $C_{4}$ are obtained, the four dimensionless constants $\alpha_3$, $\alpha_4$, $\beta_3$, and $\beta_4$, are determined from Eqs.\ (\ref{eq:C3}) and (\ref{eq:C4}).
(For the $\Lambda(t)$ type, we have $\beta_{3}=(2/3)C_{3}$ and $\beta_{4}=(2/3)C_{4}$, whereas we have $\beta_{3}=C_{3}$ and $\beta_{4}=C_{4}$ for the BV type.)
The luminosity distance \cite{Sato1} is generally given by 
\begin{equation}
  \left ( \frac{ H_{0} }{ c } \right )   d_{L}  
      =   (1+z)  \int_{1}^{1+z}  \frac{dy} { F(y) }    , 
\label{eq:dL_00}  
\end{equation}
where the integrating variable $y$, the function $F(y)$, and the redshift $z$ are given by 
\begin{equation}
  y = \frac{a_0} {a},  \hspace{2mm} F(y)  = \frac{ H }{ H_{0} } , \hspace{2mm} \textrm{and} \hspace{2mm}    z \equiv \frac{ a_0 }{ a } -1   . 
\end{equation}
Substituting Eq.\ (\ref{eq:H/H0_(C1C3)_00}) into Eq.\ (\ref{eq:dL_00}), we obtain the luminosity distance for the $H$ version.
For the standard $\Lambda$CDM model, the luminosity distance of a spatially flat universe is given as 
\begin{align}
 \left ( \frac{H_0}{c} \right )   d_{L} & =  (1+ z) \int_{0}^{z} dx [ (1+x)^2 (1+ \Omega_{m} x)  \notag  \\
                                                   & \quad    -x(2+x) \Omega_{\Lambda}]^{-1/2}   , 
\label{eq:dL(CDM)}
\end{align}
where $\Omega_{m}$ and $\Omega_{\Lambda}$ represent the density parameters for matter and $\Lambda$, respectively. 
In the standard $\Lambda$CDM model, $\Omega_{m}$ and $\Omega_{\Lambda}$ are given by $\Omega_{m} = \frac{\rho_{m}}{\rho_{c0}} = \frac{8\pi G \rho_m}{3H_{0}^2}$ and $\Omega_{\Lambda}= \frac{\Lambda}{3 H_{0}^2}$ \cite{Carroll01}.
Here $\rho_{m}$ is the density of matter, including baryonic and dark matter.
In the present paper, we consider a spatially flat universe given by $\Omega_{\textrm{total}}=  \Omega_{m} + \Omega_{\Lambda} = 1$, neglecting the density parameter $\Omega_{r}$ for the radiation \cite{Koma4,Koma4bc,Koma5}.
In particular, we consider the universe in which $(\Omega_{m}, \Omega_{\Lambda}) = (0.315, 0.685)$.
This universe is obtained from a fine-tuned standard $\Lambda$CDM model, which takes into account the recent Planck 2013 best fit values \cite{Planck2013}. 

As examined in Ref.\ \cite{Koma5}, we determine $C_{3}$ for the $H$ version, through fitting with the fine-tuned standard $\Lambda$CDM model, minimizing the function given by
\begin{equation}
\chi^{2} (C_{3}) = \sum\limits_{i=0}^{N_{z}} { \left[ \frac{   d_{L,\Lambda} (z)  - d_{L} (z; C_{3})  }{ d_{L,\Lambda} (z) }   \right]^{2}   }   ,
\label{eq:min}
\end{equation}
where $d_{L,\Lambda} (z)$ and $d_{L} (z; C_{3})$ are the luminosity distances for the fine-tuned standard $\Lambda$CDM model and the $H$ version, respectively.
Through fitting, $C_{3}$ is approximately determined to be $0.884$ \cite{Koma5}.
The dimensionless constants for the $\Lambda(t)$-$H$ and BV-$H$ models are summarized in Table \ref{tab-parameter}.

Next, we determine the dimensionless constants for the $C_{\rm{cst}}$ version, i.e., the $\Lambda(t)$-$C_{\rm{cst}}$ and BV-$C_{\rm{cst}}$ models.
To this end, we determine $C_{4}$.
As shown in Eq.\ (\ref{eq:H/H0(C1C4)_00}), the evolution of the Hubble parameter for the $C_{\rm{cst}}$ version is given as 
\begin{equation}
 \left ( \frac{H} {H_{0}} \right )^{2}   
=  \left ( 1-  \frac{ C_{4} }{ C_{1} }  \right )   \left ( \frac{ a } {  a_{0} } \right )^{ -2 C_{1}}   +  \frac{ C_{4} }{ C_{1} }   , 
\label{eq:H/H0(C1C4)_01}
\end{equation}
where $C_{1}$ can be replaced by $C_{m}=3/2$ [Eq.\ (\ref{eq:Cm=1.5})].
Equation (\ref{eq:H/H0(C1C4)_01}) is equivalent to the solution of the standard $\Lambda$CDM model. 
This implies that the constant term $C_{4}/C_{1}$ behaves like $\Omega_{\Lambda}$ in the standard $\Lambda$CDM model.
Therefore, we determine $C_{4}$ from $ C_{1} \Omega_{\Lambda}$, without fitting.
Consequently, $C_{4}$ is determined to be $C_{1} \Omega_{\Lambda} = 1.5 \times 0.685 \simeq 1.03$. 
The dimensionless constants for the $\Lambda(t)$-$C_{\rm{cst}}$ and BV-$C_{\rm{cst}}$ models are summarized in Table \ref{tab-parameter}.

\begin{figure} [t] 
\begin{minipage}{0.495\textwidth}
\begin{center}
\scalebox{0.3}{\includegraphics{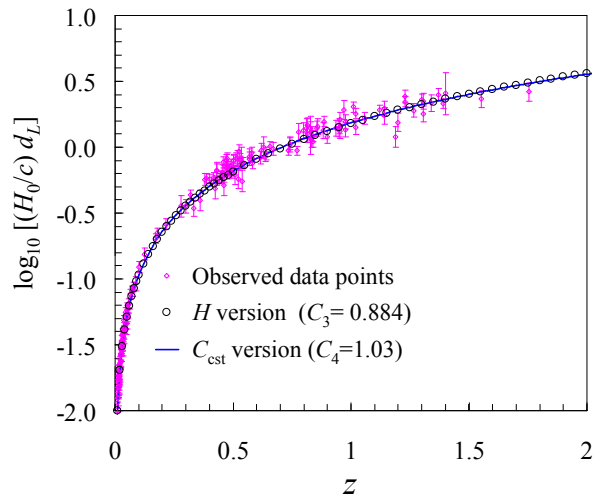}}
\end{center}
\end{minipage}
\caption{ (Color online). Dependence of luminosity distance $d_L$ on redshift $z$.
The $H$ version corresponds to the $\Lambda(t)$-$H$ and BV-$H$ models, whereas the $C_{\rm{cst}}$ version corresponds to the $\Lambda(t)$-$C_{\rm{cst}}$ and BV-$C_{\rm{cst}}$ models. 
The open diamonds with error bars are supernova data points taken from Ref.\ \cite{Riess2007SN1}. 
For the supernova data points, $H_{0}$ is set to $67.3$ km/s/Mpc \cite{Planck2013}.
Dimensionless constants for the $H$ and $C_{\rm{cst}}$ versions are summarized in Table \ref{tab-parameter}.  
Note that $d_L$ for the $H$ version is equivalent to $d_L$ examined in our previous work \cite{Koma5}.  }
\label{Fig-dL-z}
\end{figure}

\begin{figure} [t]  
\begin{minipage}{0.495\textwidth}
\begin{center}
\scalebox{0.3}{\includegraphics{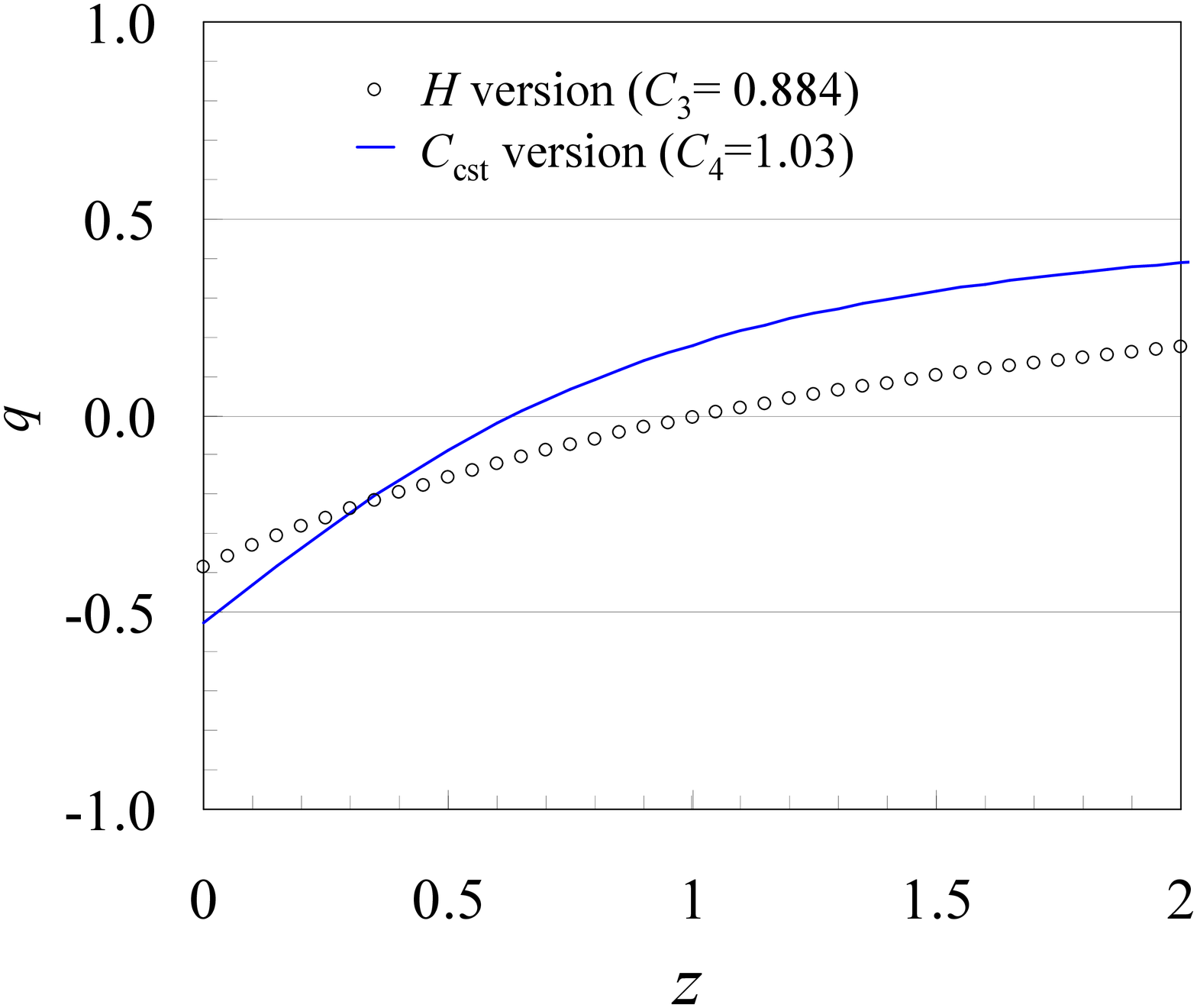}}
\end{center}
\end{minipage}
\caption{ (Color online). Dependence of temporal deceleration parameter $q$ on redshift $z$.
The $H$ version corresponds to the $\Lambda(t)$-$H$ and BV-$H$ models, whereas the $C_{\rm{cst}}$ version corresponds to the $\Lambda(t)$-$C_{\rm{cst}}$ and BV-$C_{\rm{cst}}$ models. }
\label{Fig-q-z}
\end{figure}

To observe the properties of the $H$ and $C_{\rm{cst}}$ versions, the luminosity distance $d_L$ is shown in Fig.\ \ref{Fig-dL-z}.
The $H$ version corresponds to the $\Lambda(t)$-$H$ and BV-$H$ models, whereas the $C_{\rm{cst}}$ version corresponds to the $\Lambda(t)$-$C_{\rm{cst}}$ and BV-$C_{\rm{cst}}$ models. 
As shown in Fig.\ \ref{Fig-dL-z}, both the $H$ and $C_{\rm{cst}}$ versions agree well with the supernova data points.
This is because, for the $H$ version, $C_{3}=0.884$ is determined through fitting with a fine-tuned standard $\Lambda$CDM model.
(The background evolution of the universe in the $C_{\rm{cst}}$ version is the same as that in the fine-tuned standard $\Lambda$CDM model.)  
In addition, as discussed in Ref.\ \cite{Koma5}, the $H$ and $C_{\rm{cst}}$ versions can describe a decelerating and accelerating universe.
To confirm this, we examine a temporal deceleration parameter $q$ defined by 
\begin{equation}
q \equiv  - \left ( \frac{\ddot{a} } {a H^{2}} \right )  , 
\label{eq:q_def}
\end{equation}
where positive $q$ represents deceleration and negative $q$ represents acceleration.
($\ddot{a}/a$ is equal to $\dot{H} + H^{2}$.) 
Substituting Eq.\ (\ref{eq:dHC1C3hC4h(H)}) or Eq.\ (\ref{eq:dHC1C3hC4h(Cst)}) into Eq.\ (\ref{eq:q_def}), and using $\hat{C}_{3}=C_{3} H_{0}$ or $\hat{C}_{4}=C_{4} H_{0}^{2}$, 
we obtain $q$ for the $H$ version or the $C_{\rm{cst}}$ version, respectively, given as
\begin{equation} 
 q = 
    \begin{cases} 
         C_{1}    -  \frac{ C_{3} }{ H/H_{0} } - 1                &   (H                \hspace{1mm} \rm{version}) ,         \\
         C_{1}    -  \frac{ C_{4} }{ (H/H_{0})^{2} } - 1        &   (C_{\rm{cst}} \hspace{1mm} \rm{version}) .         \\
    \end{cases}
\label{eq:q_C3C4}
\end{equation}
From Eq.\ (\ref{eq:q_C3C4}), we can calculate each temporal deceleration parameter.
As shown in Fig.\ \ref{Fig-q-z}, both the $H$ and $C_{\rm{cst}}$ versions describe a decelerated and accelerated expansion of the universe in low redshift.
In this way, the $H$ version is similar to the $C_{\rm{cst}}$ version, when we focus on the background evolution of the universe.
Note that the original entropic-force model \cite{Easson1,Easson2} cannot describe a decelerating and accelerating universe \cite{Basilakos1,Sola_2013a}, because $H$ and constant entropic-force terms are not included.

\begin{figure} [t] 
\begin{minipage}{0.495\textwidth}
\begin{center}
\scalebox{0.3}{\includegraphics{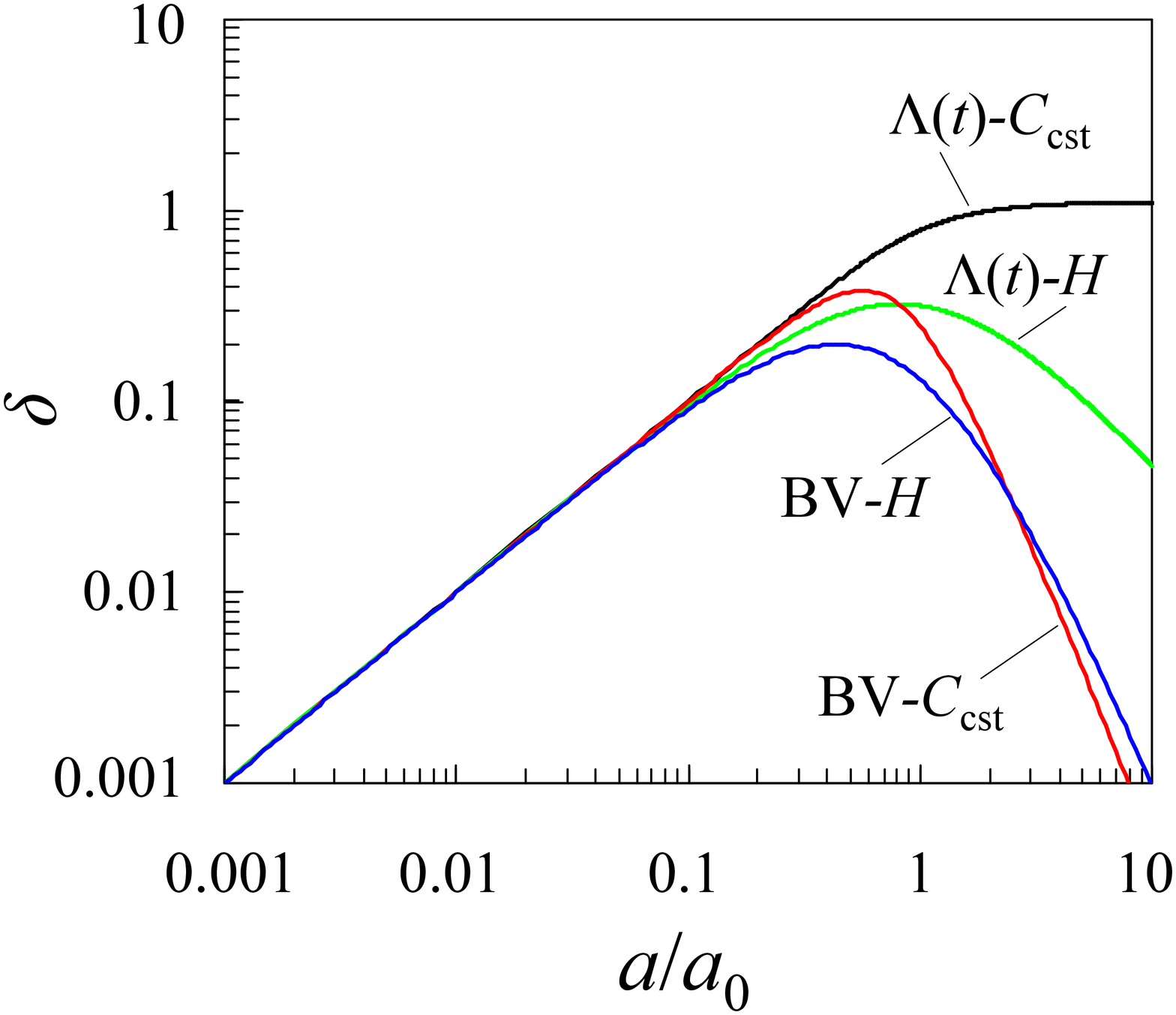}}
\end{center}
\end{minipage}
\caption{ (Color online). Evolution of density perturbation growth factor $\delta$ in the four entropic-force models.  
Note that $\delta$ for the $\Lambda (t)$-$H$ model is normalized (see the text.). }
\label{Fig-delta-a}
\end{figure}

We now examine first-order density perturbations in the four models.
To this end, we observe the evolution of the perturbation growth factor $\delta$.
As described in the previous section, $\delta$ for the BV-$H$ and BV-$C_{\rm{cst}}$ models is numerically solved by using the initial conditions of the Einstein--de Sitter growing model, i.e., $\delta (\tilde{a}_{i}) = \tilde{a}_{i}$ and $\delta^{\prime}  (\tilde{a}_{i}) = \tilde{a}_{i}$, where $\tilde{a}_{i} = a_{i}/a_{0} = 10^{-3}$.
In contrast, we calculate $\delta$ for the $\Lambda (t)$-$H$ and $\Lambda (t)$-$C_{\rm{cst}}$ models without using the initial conditions.
Consequently, we find that $\delta (\tilde{a}_{i}) $ for the $\Lambda (t)$-$H$ model is slightly smaller than $10^{-3}$. 
Therefore, $\delta$ for the $\Lambda (t)$-$H$ model is normalized so that $\delta (\tilde{a}_{i}) = \tilde{a}_{i} = 10^{-3}$ is satisfied.  
The normalized values are plotted in Fig.\ \ref{Fig-delta-a}.
The normalization for the $\Lambda (t)$-$H$ model does not influence the following discussion.
(The $\Lambda (t)$-$C_{\rm{cst}}$ model satisfies $\delta (\tilde{a}_{i}) = \tilde{a}_{i} = 10^{-3}$ without normalization.)

For small $a/a_{0}$ ($a/a_{0} \lessapprox 0.1$), $\delta$ increases with $a/a_{0}$, as shown in Fig.\ \ref{Fig-delta-a}. 
Thereafter, the increase of $\delta$ for the $\Lambda(t)$-$H$, BV-$H$, and BV-$C_{\rm{cst}}$ models tends to gradually slow. 
For $a/a_{0} \gtrapprox 1$, $\delta$ for the three models decreases, whereas $\delta$ for the $\Lambda(t)$-$C_{\rm{cst}}$ does not decrease.
It is clearly shown that density perturbations for the $\Lambda(t)$-$C_{\rm{cst}}$ and BV-$C_{\rm{cst}}$ models are different from each other.
However, as mentioned previously, the background evolution of the universe in the BV-$C_{\rm{cst}}$ model is the same as that for the $\Lambda(t)$-$C_{\rm{cst}}$ model.
Similarly, density perturbations for the $\Lambda(t)$-$H$ and BV-$H$ models are different from each other although the background evolution in the two models is the same.
In addition, for $a/a_{0} \gtrapprox 0.5$, $\delta$ for the BV types decreases significantly, in comparison with the $\Lambda(t)$ types.

\begin{figure} [t] 
\begin{minipage}{0.495\textwidth}
\begin{center}
\scalebox{0.31}{\includegraphics{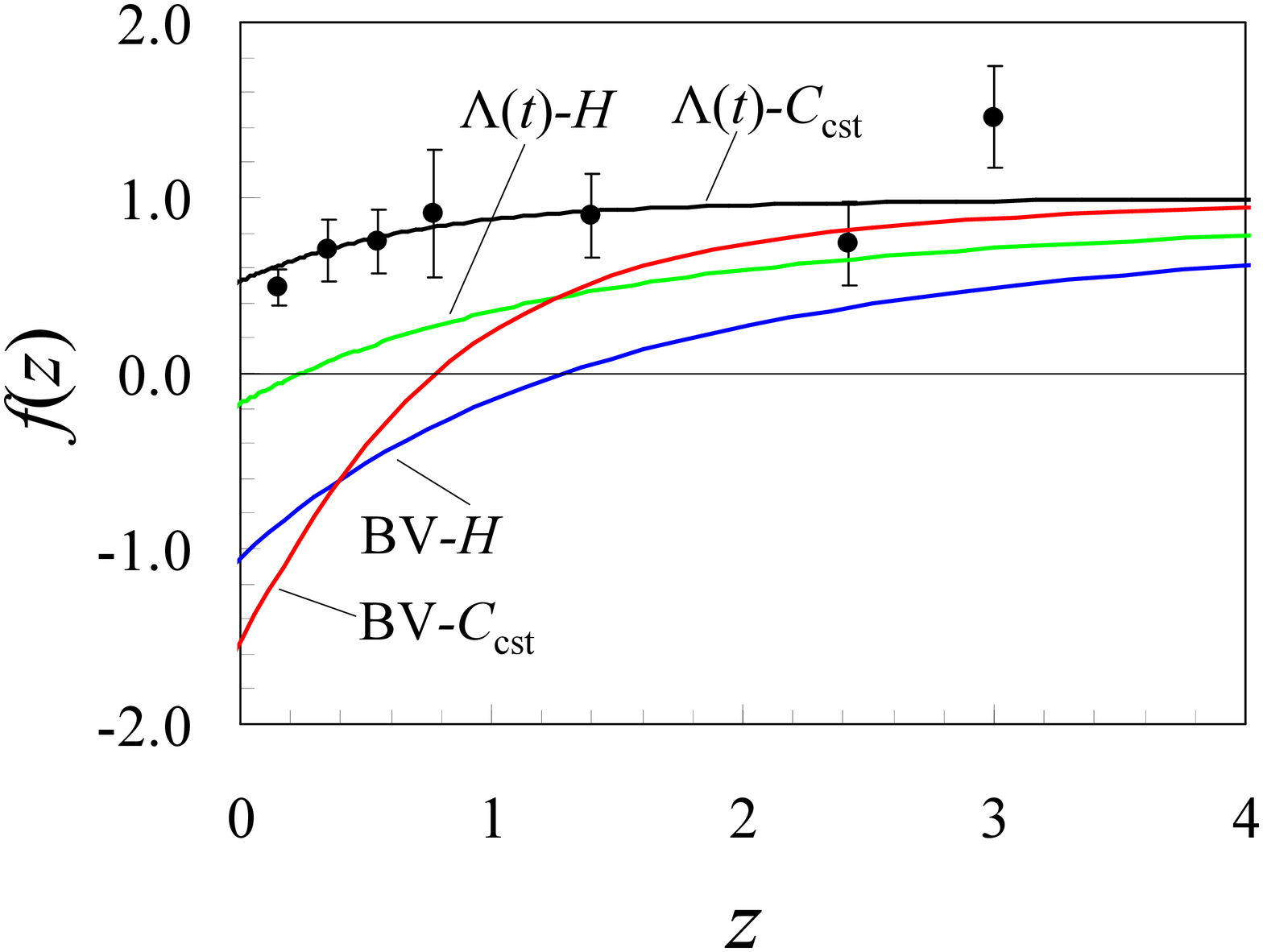}}
\end{center}
\end{minipage}
\caption{ (Color online). Evolution of the growth rate $f(z)$ of clustering in the four entropic-force models.
The closed circles with error bars are the observed data points taken from a summary of Ref.\ \cite{Lima2011}.  }
\label{Fig-f(z)-z}
\end{figure}

Finally, we observe the evolution of an indicator of clustering, namely, the growth rate of clustering \cite{Peebles_1993}.
(For the $\Lambda(t)$CDM and CCDM models, the growth rate has been closely examined, e.g., see Refs.\ \cite{Sola_2009,Lima2011}.) 
The growth rate $f(z)$ of clustering is calculated as
\begin{equation}
 f(z) = \frac{d \ln \delta }{  d \ln a } = - (1 + z ) \frac{d \ln \delta }{  dz }     .
\label{eq:f(z)}
\end{equation}
The evolution of the growth rate of clustering in the four models is shown in Fig.\ \ref{Fig-f(z)-z}.
The observed data points are taken from a summary of Ref.\ \cite{Lima2011}.
Note that each original data point is given in Refs.\ \cite{Colless2001,Guzzo2008,Tegmark2006,Ross2007,Angela2007,Viel2004,McDonald2005}.
As shown in Fig.\ \ref{Fig-f(z)-z}, for high $z$ ($z \gtrapprox 2$), the growth rate $f(z)$ of each model is positive and is likely consistent with the observed data points.
For low $z$ ($z \lessapprox 1$), $f(z)$ for the $\Lambda(t)$-$H$, BV-$H$, and BV-$C_{\rm{cst}}$ models tends to be negative and disagrees with the observed growth rate. 
This is because, as shown in Fig.\ \ref{Fig-delta-a}, $\delta$ for the three models decays at high $a/a_{0}$ (corresponding to low $z$). 
As shown in Fig.\ \ref{Fig-q-z}, the $H$ version (i.e., the $\Lambda(t)$-$H$ and BV-$H$ models) can describe a decelerating and accelerating universe predicted by the standard $\Lambda$CDM model. 
However, the $\Lambda(t)$-$H$ and BV-$H$ models disagree with the observed growth rate of clustering [Fig.\ \ref{Fig-f(z)-z}].
In addition, the BV-$C_{\rm{cst}}$ model disagrees with the observed growth rate, 
although its background evolution is the same as the $\Lambda(t)$-$C_{\rm{cst}}$ model.
Of course, as examined in Ref.\ \cite{Lima2011}, the BV-$C_{\rm{cst}}$ model agrees with the observed growth rate 
if $c_{\rm{eff}}^{2}$ is set to be $-1$ assuming that $c_{\rm{eff}}^{2}$ is a free parameter.
However, in the present paper, we do not consider the case for $c_{\rm{eff}}^{2} \neq 0$, as discussed in Sec.\ \ref{BV-Cst model}. 

In contrast, $f(z)$ for the $\Lambda(t)$-$C_{\rm{cst}}$ model agrees well with the observed growth rate, even for low $z$, as shown in Fig.\ \ref{Fig-f(z)-z}. 
This is because the formulation of the $\Lambda(t)$-$C_{\rm{cst}}$ model is equivalent to that of the standard $\Lambda$CDM model.
We find that the $\Lambda(t)$ types (i.e., the $\Lambda(t)$-$H$ and $\Lambda(t)$-$C_{\rm{cst}}$ models) are consistent with the observed growth rate, in comparison with the examined BV types (i.e., the BV-$H$ and BV-$C_{\rm{cst}}$ models).
This indicates that the $\Lambda(t)$ type, especially the $\Lambda(t)$-$C_{\rm{cst}}$ model, is suitable for describing structure formations.
Note that an entropy $S_{r4}$ proportional to $r_{H}^{4}$ is required for the $\Lambda(t)$-$C_{\rm{cst}}$ model, as discussed in Appendix \ref{Entropic-force from the hyper-dimension entropy}.

In the present paper, we study the evolution of the universe in the four entropic-force models, obtained by combining the $H$ and constant entropic-force terms with the $\Lambda(t)$ and BV types.
Similar results have been examined in bulk viscous models, CCDM models, and $\Lambda(t)$CDM models.
For example, cosmological models similar to the BV-$H$, BV-$C_{\rm{cst}}$, and $\Lambda(t)$-$H$ models have been closely investigated in Refs.\ \cite{Barrow21}, \cite{Lima2011}, and \cite{Sola_2009}, respectively.
Our results are consistent with those examined in the previous works.

\section{Conclusions}
\label{Conclusions}

Entropic-force models are categorized into two types.
The first is the $\Lambda(t)$ type similar to $\Lambda(t)$CDM models, 
and the second is the BV type similar to bulk viscous models (and CCDM models).
In order to examine the two types systematically, we have considered an extended entropic-force model which includes $H$ and constant $C_{\rm{cst}}$ terms.
In particular, we have focused on the $H$ and $C_{\rm{cst}}$ terms separately, in a homogeneous, isotropic, and spatially flat matter-dominated universe.
The constant entropic-force term considered here is derived from an entropy $S_{r4}$ proportional to $r_{H}^{4}$, assuming $S_{r4}$ as one of the possible entropies. 

In the present paper, we have examined four models, the $\Lambda(t)$-$H$, $\Lambda(t)$-$C_{\rm{cst}}$, BV-$H$, and BV-$C_{\rm{cst}}$ models, which are obtained from combining the $H$ and $C_{\rm{cst}}$ terms with the $\Lambda(t)$ and BV types.
The four models agree well with observed supernova data points and describe a decelerated and accelerated expansion of the universe.
In order to examine first-order density perturbations in the four models, we used two formulations proposed by Basilakos \textit{et al.} \cite{Sola_2009} and Jesus \textit{et al.} \cite{Lima2011}.
The two formulations can be summarized using a neo-Newtonian approach and, therefore, we have proposed a unified formulation which helps to observe the two formulations systematically.
In addition, we have extended the formulation, in order to study the BV-$H$ model.
Consequently, for large $a/a_{0}$, the perturbation growth factor $\delta$ for the $\Lambda(t)$-$H$, BV-$H$, and BV-$C_{\rm{cst}}$ models decreases, whereas $\delta$ for $\Lambda(t)$-$C_{\rm{cst}}$ does not decrease.
Therefore, for low redshift, the growth rate for the $\Lambda(t)$-$H$, BV-$H$, and BV-$C_{\rm{cst}}$ models tends to be negative and disagrees with the observed growth rate.
It is found that for low redshift, $H$ versions (i.e., the $\Lambda(t)$-$H$ and BV-$H$ models) are not consistent with structure formations, though the $H$ version describes a decelerating and accelerating universe. 
In contrast, the growth rate for the $\Lambda(t)$-$C_{\rm{cst}}$ model agrees well with the observed growth rate. 
Interestingly, the BV-$C_{\rm{cst}}$ model disagrees with the observed growth rate, 
although its background evolution of the universe is the same as that of the $\Lambda(t)$-$C_{\rm{cst}}$ model.
(Note that we have assumed $c_{\rm{eff}}^{2} = 0$ in the present study. 
The BV-$C_{\rm{cst}}$ model agrees with the observed growth rate if $c_{\rm{eff}}^{2}$ is considered to be a free parameter, as examined by Jesus \textit{et al.} \cite{Lima2011}.)

It is also found that $\Lambda(t)$ types (the $\Lambda(t)$-$H$ and $\Lambda(t)$-$C_{\rm{cst}}$ models) are consistent with the observed growth rate, in contrast with BV types (the BV-$H$ and BV-$C_{\rm{cst}}$ models). 
Therefore, in entropic-force models, the $\Lambda(t)$ type is likely suitable for describing density perturbations or structure formations.
Through the present study, we have revealed fundamental properties of the two types of entropic-force models systematically.
Of course, similar cosmological models have been discussed in bulk viscous models \cite{Barrow21}, CCDM models \cite{Lima2011}, and $\Lambda(t)$CDM models \cite{Sola_2009}.
Our results are related to those cosmological models although the theoretical backgrounds are different.

\appendix

\section{Derivation of entropic-force terms} 
\label{Derivation of entropic-force}

In the entropic cosmology suggested by Easson \textit{et al.} \cite{Easson1}, the horizon of the universe is assumed to have an associated entropy and an approximate temperature.
In this paper, we use the Hubble horizon as the preferred screen, because the apparent horizon coincides with the Hubble horizon in a spatially flat universe \cite{Easson1}.
(If we consider a spatially non-flat universe, we would use the apparent horizon as the preferred screen rather than the Hubble horizon.) 
The Hubble horizon (radius) $r_{H}$ is given by
\begin{equation}
     r_{H} = \frac{c}{H}   .
\label{eq:rH}
\end{equation}
The temperature $T$ on the Hubble horizon is given by 
\begin{equation}
 T = \frac{ \hbar H}{   2 \pi  k_{B}  } \times   \gamma  =  \frac{ \hbar }{   2 \pi  k_{B}  }  \frac{c}{ r_{H} }   \gamma    ,   
\label{eq:T0}
\end{equation}
where $k_{B}$ and $\hbar$ are the Boltzmann constant and the reduced Planck constant, respectively.
The reduced Planck constant is defined by $\hbar \equiv h/(2 \pi)$, where $h$ is the Planck constant.
As described in Refs.\ \cite{Koma4,Koma4bc,Koma5}, the temperature considered here is obtained by multiplying the horizon temperature, $ \hbar H /( 2 \pi k_{B} ) $, by $\gamma$, 
a non-negative free parameter on the order of $O(1)$.
(A similar parameter for the screen temperature has been discussed in Refs.\ \cite{Easson1,Cai1_Cai2_Qiu1}.)
In the present study, we use the temperature on the horizon, assuming thermal equilibrium states based on a single holographic screen \cite{Easson1,Easson2}. 

In the following, we discuss three entropic-force terms, the  $H^{2}$, $H$, and constant terms, which are derived from an area entropy $S_{r2}$, a volume entropy $S_{r3}$, and an entropy $S_{r4}$ proportional to $r_{H}^{4}$, respectively.
In Secs.\ \ref{Entropic-force from the area entropy} and \ref{Entropic-force from the volume entropy}, we derive the $H^{2}$ and $H$ terms, according to the works of Easson {\it et al.} \cite{Easson1} and the present authors \cite{Koma5}.
In Sec.\ \ref{Entropic-force from the hyper-dimension entropy}, we derive the constant  entropic-force term from $S_{r4}$ proportional to $r_{H}^{4}$.

\subsection{$H^{2}$ terms derived from the area entropy $S_{r2}$} 
\label{Entropic-force from the area entropy}

In the original entropic-force model \cite{Easson1,Easson2}, an associated entropy on the Hubble horizon is given as
\begin{equation} 
S_{r2}  = \frac{ k_{B} c^3 }{  \hbar G }  \frac{A_{H}}{4}   ,
\label{eq:SH(r2)}
\end{equation}
where $A_{H}$ is the surface area of a sphere with the Hubble radius $r_{H}$.
This is the Bekenstein entropy (area entropy) which is proportional to $A_{H}$ and $r_{H}^2$ \cite{Koma4,Koma4bc,Koma5}.
Substituting $A_{H}=4 \pi r_{H}^2 $ into Eq.\ (\ref{eq:SH(r2)}), and using $r_{H}= c/H$,  we have 
\begin{equation}
S_{r2}
 = \frac{ k_{B} c^3 }{  \hbar G }         \frac{A_{H}}{4}       
 =  \left ( \frac{ \pi k_{B} c^5 }{ \hbar G } \right )  \frac{1}{H^2}  =  K  \frac{1}{H^2}   ,   
\label{eq:SH(r2)2}      
\end{equation}
where $K$ is a positive constant \cite{Koma4,Koma5} given by
\begin{equation}
  K =  \frac{  \pi  k_{B}  c^5 }{ \hbar G }     .
\label{eq:K-def}
\end{equation}
The entropic-force $F$ can be given by 
\begin{equation}
    F  =  -  \frac{dE}{dr}  =  - T \frac{dS}{dr}   \left ( =    - T \frac{dS}{dr_{H}} \right )        ,     
\label{eq:F}
\end{equation}
where the minus sign indicates the direction of increasing entropy or the screen corresponding to the horizon \cite{Easson1}.
The entropic-force $F_{r2}$ derived from the area entropy $S_{r2}$ is given as
\begin{equation}
    F_{r2}  =    - T \frac{dS_{r2}}{dr_{H}}  =  - \gamma  \frac{c^{4}}{G}   .
\label{eq:F(r2)}
\end{equation}
Therefore, the pressure $p_{r2}$ \cite{Easson1} is given by
\begin{equation}
 p_{r2} =    \frac{ F_{r2} } {A_{H}}   
          =     -  \gamma  \frac{c^{4}}{G}    \frac{1} {4 \pi (c/H)^2} =  -  \gamma  \frac{c^{2}}{4 \pi G}  H^{2}           .
\label{eq:P(r2)}
\end{equation}
Since Eq.\ (\ref{eq:P(r2)}) indicates negative pressure, the entropic-force model can explain an accelerated expansion of the late universe \cite{Easson1}.
The pressure $p_{r2}$ is proportional to $H^{2}$ which corresponds to entropic-force terms.
In Refs.\ \cite{Easson1,Koma5}, the acceleration equation is given as
\begin{equation}
  \frac{ \ddot{a} }{ a }  =   -  \frac{ 4\pi G }{ 3 }  \left (  \rho +  \frac{ 3p }{c^2}  \right )    +   \gamma  H^{2}    .
\label{eq:FRW2_(r2)}
\end{equation}
The last term, i.e., the $\gamma H^2$ term, is the so-called entropic-force term derived from the area entropy $S_{r2}$.

\subsection{$H$ terms derived from the volume entropy $S_{r3}$} 
\label{Entropic-force from the volume entropy}

In this subsection, instead of area entropy, we consider a volume entropy \cite{Koma5}.
Recently, Tsallis and Cirto have suggested a generalized black-hole entropy using appropriate nonadditive generalizations for $d$-dimensional systems \cite{Tsallis2012}.
In their study, a nonadditive entropy (for a set of $W$ discrete states) is defined by
\begin{equation}
S_{\delta_g}  =  k_{B} \sum_{i=1}^{W}  p_{i} \left ( \ln \frac{1}{{p_{i}}  } \right ) ^{\delta_g}  \quad (\delta_g >0)   ,
\end{equation}
where $p_{i}$ is a probability distribution \cite{Tsallis2012}. 
(For other nonextensive entropies, e.g., Tsallis' entropy \cite{Tsa0}, see Ref.\ \cite{Tsa1}.)
When $\delta_g = 1$, $S_{\delta_g}$ recovers the Boltzmann--Gibbs entropy.
Tsallis and Cirto demonstrated that a generalized black-hole entropy can be written as
\begin{equation}
 \frac{ S_{\delta_g = 3/2 } } { k_{B} }  \propto \left (  \frac{ S_{\rm{B}}  } { k_{B} } \right )   ^{\frac{3}{2}}   , 
\label{eq:S-SBH}
\end{equation}
where the event horizon area of a black hole is used for the Bekenstein black-hole entropy $S_{\rm{B}}$.
As examined in our previous study \cite{Koma5}, we apply this entropy to an entropy for entropic cosmology.
Using $A_{H}=4 \pi r_{H}^2$, we have the entropy $S$ on the Hubble horizon evaluated as 
\begin{equation}
  S  \propto   A_{H}^{\frac{3}{2}}  \propto   r_{H}^{3}  .
\label{eq:Sprop}
\end{equation}
Accordingly, we assume the volume entropy $S_{r3}$ given by 
\begin{equation}
S_{r3}  =    \frac{  \pi  k_{B} c^3 }{  \hbar G } \times  \zeta r_{H}^{3}  ,
\label{eq:S(r3)}
\end{equation}
where $\zeta$ is a non-negative free-parameter and is a dimensional constant \cite{Koma5}.
Substituting $r_{H}= c/H$ into Eq.\ (\ref{eq:S(r3)}), we obtain
\begin{equation}
S_{r3}  = \frac{  \pi  k_{B} c^3 }{  \hbar G } \times  \zeta  \left ( \frac{c}{H} \right )^{3}   =  K  c \zeta  \frac{1}{H^3}   , 
\label{eq:S(r3)2}      
\end{equation}
where $K$ is $ \pi  k_{B}  c^5  / (\hbar G) $ given by Eq.\ (\ref{eq:K-def}).
Therefore, the entropic-force $F_{r3}$ derived from the volume entropy $S_{r3}$ \cite{Koma5} is given as
\begin{equation}
    F_{r3}   =    - T \frac{dS_{r3}}{dr_{H}}      =  -  \gamma  \frac{c^{4}}{G} \left ( \frac{3 c \zeta}{2} \frac{1}{H} \right )    .
\label{eq:F(r3)}
\end{equation}
The pressure $p_{r3}$ is given as 
\begin{equation}
   p_{r3}  =    \frac{ F_{r3} } {A_{H}}    =  -  \gamma  \frac{c^{2}}{4 \pi G}   \frac{3 c \zeta }{2} H                         .
\label{eq:P(r3)}
\end{equation}
The obtained entropic-force term is an $H$ term, unlike in the case of the original entropic-force model.
That is, the $H$ term is derived from the volume entropy $S_{r3}$.
As examined in Ref.\ \cite{Koma5}, the entropic-force model which includes the $H$ term describes a decelerating and accelerating universe.

\subsection{Constant terms derived from an entropy $S_{r4}$ proportional to $r_{H}^{4}$} 
\label{Entropic-force from the hyper-dimension entropy}

In this subsection, we assume an entropy $S_{r4}$ proportional to $r_{H}^4$, which is defined by
\begin{equation}
S_{r4} =    \frac{  \pi  k_{B} c^3 }{  \hbar G } \times  \psi r_{H}^{4}           ,
\label{eq:S(r4)}
\end{equation}
where $\psi$ is a non-negative free-parameter and a dimensional constant.
Of course, the origin of such an entropy is not clear. 
However, it is possible to consider $S_{r4}$ as a possible entropy if extra dimensions were assumed. 
Substituting $r_{H}= c/H$ into Eq.\ (\ref{eq:S(r4)}), we obtain
\begin{equation}
S_{r4}  = \frac{  \pi  k_{B} c^3 }{  \hbar G } \times  \psi r_{H}^{4} =  K  c^{2} \psi  \frac{1}{H^4}   ,  
\label{eq:S(r4)2}      
\end{equation}
where $K$ is $ \pi  k_{B}  c^5  / (\hbar G) $ [Eq.\ (\ref{eq:K-def})].
Substituting Eqs.\ (\ref{eq:T0}) and (\ref{eq:S(r4)}) into Eq.\ (\ref{eq:F}), and using $r_{H}= c/H$, we have the entropic-force $F_{r4}$ given as
\begin{align}
    F_{r4}  &=      - T \frac{ dS_{r4} }{dr_{H}}           
                =  -  \frac{ \hbar }{ 2 \pi k_{B}  }  \frac{c}{ r_{H} }  \gamma   \times   \frac{d}{dr_{H}}  \left [ \frac{  \pi  k_{B} c^3 }{  \hbar G } \times  \psi r_{H}^{4}    \right ]     \notag \\
           &=  -  \gamma  \frac{c^{4}}{G} \left ( 2 \psi r_{H}^{2} \right ) =  -  \gamma  \frac{c^{4}}{G} \left ( 2 c^{2} \psi \frac{1}{H^{2}} \right )    .
\label{eq:F(r4)}
\end{align}
From Eq.\ (\ref{eq:F(r4)}), the pressure $p_{r4}$ derived from $S_{r4}$ is given as 
\begin{align}
   p_{r4}  &=    \frac{ F_{r4} } {A_{H}}   =  -  \gamma  \frac{c^{4}}{G} \left ( 2 c^{2} \psi \frac{1}{H^{2}} \right )   \frac{1} {4 \pi r_{H}^2}     \notag \\ 
                     &=  -  \gamma  \frac{c^{4}}{G} \left ( 2 c^{2} \psi \frac{1}{H^{2}} \right )  \frac{1}{4 \pi (c/H)^2}                                  
                       =  -  \gamma  \frac{c^{2}}{4 \pi G}   (2 c^{2} \psi)                        .
\label{eq:P(r4)}
\end{align}
The constant term (similar to a cosmological constant) is derived from $S_{r4}$ proportional to $r_{H}^{4}$.
While the origin of $S_{r4}$ is not clear and it is
therefore important to ultimately clarify the origin of $S_{r4}$, 
we do not discuss this in the present study, though we assume $S_{r4}$ as a possible model.
Note that a similar constant term can be obtained if a bulk viscosity $\xi$ of cosmological fluids is given by $\xi \propto 1/ H$ ($\sim 1/ T$).

\section{Background evolution in an extended entropic-force model}
\label{Solutions of an extended entropic-force model}

We review the background evolution of the universe in an extended entropic-force model given by Eq.\ (\ref{eq:dHC1C3hC4h}).
In fact, Eq.\ (\ref{eq:dHC1C3hC4h}) is essentially the same as the equation for a general $\Lambda(t)$CDM model examined in Ref.\ \cite{Sola_2009}.
From Eq. (\ref{eq:dHC1C3hC4h}), we have 
\begin{equation}
 \int_{+ \infty}^{H} \frac{dy} { -  C_{1} y^2    +   \hat{C}_{3} y   +  \hat{C}_{4}}  = t   ,
\label{eq:Int_dHC1C3hC4h}
\end{equation}
where we consider $C_{1} > 0$, $\hat{C}_{3} \geq 0$, and $\hat{C}_{4} \geq 0$, except the case for $\hat{C}_{3} = \hat{C}_{4} = 0$.
Using $\hat{C}_{3} = C_{3} H_{0}$ and $\hat{C}_{4} = C_{4} H_{0}^{2}$ [Eqs.\ (\ref{eq:C3}) and (\ref{eq:C4})], and rearranging, we obtain 
\begin{equation}
 \frac{H} {H_{0}} =  \frac { (C_{3} + A) \exp[A H_{0} t] - C_{3} + A } { 2 C_{1}  (\exp[A H_{0} t] - 1 ) }           ,
\label{eq:H/H0_(g)}
\end{equation}
\begin{equation}
 \frac{a} {a_{0}} =  \frac { \left (   \exp[A H_{0} t] - 1  \right )^{ \frac{1}{C_{1}} } \exp \left [ \frac{ C_{3} - A }{ 2 C_{1} }\times H_{0} t \right ]   } 
                                   { \left (   \exp[A H_{0} t_{0}] - 1  \right )^{ \frac{1}{C_{1}} } \exp \left [ \frac{ C_{3} - A }{ 2 C_{1} }\times H_{0} t_{0} \right ]   }    ,
\label{eq:a/a0_(g)}
\end{equation}
where $A$ and $H_{0} t_{0}$ are given by 
\begin{equation}
   A = \sqrt{C_{3}^{2} + 4 C_{1} C_{4} }   ,
\label{eq:A_g}
\end{equation}
\begin{equation}
      H_{0} t_{0}  = \ln  \left [   \frac{2 C_{1} - C_{3} + A }{ 2 C_{1} - C_{3} - A }  \right ] ^{\frac{1}{A}}    .
\label{eq:H0t0_g}
\end{equation}
We can apply the above solutions to both the $\Lambda(t)$ and BV types.
The solutions are equivalent to those for the general $\Lambda(t)$CDM model  \cite{Sola_2009}.

\section{$\boldsymbol{C_{\rm{cst}}}$ version}
\label{Solutions for the model with constant terms}

We examine solutions of an entropic-force model which includes constant terms, i.e., the ${C_{\rm{cst}}}$ version. 
To this end, we neglect $H$ terms from the extended entropic-force model given by Eqs.\ (\ref{eq:f(g)}) and (\ref{eq:g(g)}).
That is, we assume $\hat{\alpha}_{3} = \hat{\beta}_{3} =0$, and therefore $\hat{C}_{3}$ of Eq.\ (\ref{eq:dHC1C3hC4h}) is $0$.
In addition, we assume that $C_{1}$ is a positive constant, in a single-fluid-dominated universe.
In the following, we extend our solution method discussed in Refs.\ \cite{Koma4,Koma5}, focusing on the background evolution of the universe.

When $\hat{C}_{3}=0$, we can rearrange Eq.\ (\ref{eq:dHC1C3hC4h}) as
\begin{equation}
 \dot{H}  = \frac{ dH }{ dt }  = -  C_{1} H^2   +  \hat{C}_{4}  .
\label{eq:dHC1C4h_1}
\end{equation}
Therefore, we have 
\begin{equation}
  \frac{ d H }{ d N } =  - C_{1} H  + \frac{ \hat{C}_{4} }{H} ,
\label{eq:dHdN(r4)}
\end{equation}
where $N$ is defined by 
\begin{equation}
   N  \equiv \ln a   \quad \textrm{and therefore} \quad  dN   = \frac{da}{a}   .  
\end{equation} 
We can solve Eq.\ (\ref{eq:dHdN(r4)}) when $C_{1}$ and $\hat{C}_{4}$ are constant.
(We consider $\hat{C_{4}}$ to be a non-negative free parameter.) 
When $C_1$ and $\hat{C}_{4}$ are constant, Eq.\ (\ref{eq:dHdN(r4)}) is integrated as
\begin{equation}
 \int \frac{dH}{- C_{1} H + \frac{\hat{C}_{4}}{H} } = \int dN   .  
\end{equation}
Solving this integral, and using $N = \ln a$, we have 
\begin{equation}
C_{1} H^{2} - \hat{C}_{4}  =  D a^{- 2 C_{1}}      , 
\end{equation}
and dividing this equation by $C_{1}$ gives 
\begin{equation}
   H^{2} - \frac{\hat{C}_{4}}{C_{1}}  = \frac{D}{C_{1}} a^{- 2 C_{1}}      , 
\label{eq:Solve1(r4)}
\end{equation}
where $D$ is an integral constant.
Dividing Eq.\ (\ref{eq:Solve1(r4)}) by $ H_{0}^{2} - (\hat{C}_{4}/C_{1})  =  (D/C_{1}) a_{0}^{- 2 C_{1}}$, we have   
\begin{equation}
 \frac{ H^{2} - (\hat{C}_{4}/C_{1})  }{ H_{0}^{2} - (\hat{C}_{4}/C_{1})  }   =  \left ( \frac{ a } {  a_{0} } \right )^{ -2 C_{1}}   .
\label{eq:H/H0(r4)}
\end{equation}
Rearranging Eq.\ (\ref{eq:H/H0(r4)}) and substituting $C_{4}= \hat{C}_{4}/H_{0}^{2}$ [Eq.\ (\ref{eq:C4})] into the resulting equation, we obtain 
\begin{align}
 \left ( \frac{H} {H_{0}} \right )^{2}   &=  \left ( 1-  \frac{1}{H_{0}^{2}} \frac{ \hat{C}_{4} }{ C_{1} }  \right )   \left ( \frac{ a } {  a_{0} } \right )^{ -2 C_{1}}   +  \frac{1}{H_{0}^{2}} \frac{\hat{C}_{4}}{C_{1}}    \notag\\
                                                    &=  \left ( 1-  \frac{ C_{4} }{ C_{1} }  \right )   \left ( \frac{ a } {  a_{0} } \right )^{ -2 C_{1}}   +  \frac{ C_{4} }{ C_{1} }    , 
\label{eq:H/H0(r4)_(2)}
\end{align}
where $C_{1}$ and $C_{4}$ are determined from Eqs.\ (\ref{eq:C1}) and (\ref{eq:C4}), respectively.

Finally, we discuss the time evolution of the scale factor. 
To this end, Eq.\ (\ref{eq:H/H0(r4)_(2)}) is rearranged as
\begin{equation}
 \tilde{ H }^{2}   =   ( 1- B) \tilde{a}^{-2 C_{1}}   +  B    ,
\label{eq:H/H0(r4)_(2)AB}
\end{equation}
where $\tilde{ H }$,  $\tilde{ a }$, and $B$ are defined by 
\begin{equation}
 \tilde{ H } \equiv \frac{H}{H_0} ,  \quad  \quad  \tilde{ a } \equiv \frac{a}{a_0} ,  \quad   \quad   B \equiv  \frac{ C_{4} }{ C_{1} } .
\label{eq:def_H_a(r4)}
\end{equation}
Multiplying Eq.\ (\ref{eq:H/H0(r4)_(2)AB}) by $\tilde{ a }^{2} $, we obtain
\begin{equation}
\tilde{H}^{2} \tilde{ a }^{2}   =   \tilde{a}^{2} [( 1- B) \tilde{a}^{-2 C_{1}}  +  B]      .
\label{eq:dadt(r4)}
\end{equation}
Substituting $\tilde{H} \tilde{ a } =  (d \tilde{a} /dt)/ H_{0}$ \cite{Koma4,Koma5} into Eq.\ (\ref{eq:dadt(r4)}) and rearranging,  we have 
\begin{equation}
\frac{1}{ H_{0} } \frac{ d\tilde{a} } { dt }      =   \tilde{a} \sqrt{ ( 1- B) \tilde{a}^{-2 C_{1}}  +  B}  .
\label{eq:dadt(r4)_2}
\end{equation}
Integrating Eq.\ (\ref{eq:dadt(r4)_2}), we obtain
\begin{equation}
\int ^{\tilde{a}} _{1}  \frac{ d x  }{  x \sqrt{( 1- B) x^{  -2 C_{1}}  +  B}  }    
= \int ^{t}_{t_{0}} H_{0} dt =H_{0} (t -t_{0})   . 
\label{eq:int_a-t(r4)}
\end{equation}
Solving this integral yields 
\begin{equation}
 \frac{1}{2 \sqrt{B} C_{1}} \ln \left [ \frac{  1 + \sqrt{(1/B -1) \tilde{a}^{-2 C_{1} } +1}  }{  1 - \sqrt{(1/B -1) \tilde{a}^{-2 C_{1} } +1}  } \right ]  =   H_{0} (t -t_{0})   . 
\label{eq:2AB(r4)}
\end{equation}
Moreover, solving Eq.\ (\ref{eq:2AB(r4)}) for $\tilde{a}$ and substituting Eq.\ (\ref{eq:def_H_a(r4)}) into the resulting equation, we have   
\begin{align} 
     & \frac{a}{a_{0}}   =        \notag \\ 
     &  \left [  \frac{    \left  ( \sqrt{ \frac{C_{4}}{C_{1}} } + 1  \right   )  \exp[ 2 \sqrt{C_{4} C_{1} }  H_{0}( t - t_{0} ) ]  + \sqrt{ \frac{C_{4}}{C_{1}} } - 1     }
                                                {      2 \sqrt{ \frac{C_{4}}{C_{1}} }  \exp[  \sqrt{C_{4} C_{1} }  H_{0}( t - t_{0} ) ]                                                                                     }
                                \right   ]^{\frac{1}{C_{1}}}     ,
\label{eq:a-t(r4)}
\end{align}
and rearranging this gives   
\begin{equation} 
   \frac{a}{a_{0}}   =    \left [     \cosh  \left  (   \sqrt{C_{4} C_{1} }  t_{H_{0}} \right    )
                                          +  \sqrt{ \frac{C_{1}}{C_{4}} }  \sinh  \left (   \sqrt{C_{4} C_{1} }   t_{H_{0}}  \right    )     \right   ]^{\frac{1}{C_{1}}}     ,
\label{eq:a-t(r4)2}
\end{equation}
where $t_{H_{0}}$ is defined by
\begin{equation} 
   t_{H_{0}}  \equiv  H_{0} ( t - t_{0} )   .
\end{equation}
The equivalent equations have been extensively examined in $\Lambda$CDM and $\Lambda(t)$CDM models. 
Note that the constant term considered here is derived from an entropy $S_{r4}$ proportional to $r_{H}^4$.


\begin{thebibliography}{99}
\bibitem{C01}
The continuity equation for the $\Lambda$CDM model can be derived from the first law of thermodynamics \cite{Ryden1}, without using the Friedmann and acceleration equations.
\bibitem{C11}
An equivalent continuity equation can be obtained from the first law of thermodynamics \cite{Koma5}, using a volume entropy $S_{r3}$ of Appendix \ref{Entropic-force from the volume entropy}.
\bibitem{C12}
Using the first law of thermodynamics and $S_{r4}$ of Appendix \ref{Entropic-force from the hyper-dimension entropy}, we have an alternative continuity equation given by  
$ \dot{\rho} + 3 (\dot{a}/a) [ \rho   +  (p/c^2) ]  = -  \gamma  \frac{3}{4 \pi G} (2 c^{2} \psi )  \dot{H} / H$.
This equation agrees with Eq. (\ref{eq:drho_General0_f=g_(p=0)_L-C_Q=0}) when $\dot{H} =0$.
When $\dot{H} = - H^{2}$, the alternative continuity equation corresponds to the continuity equation for the BV-$C_{\rm{cst}}$ model [Eq.\ (\ref{eq:drho_General0_f=0_(p=0)2})] examined in Sec.\ \ref{BV-Cst model}.
\end{thebibliography}
\end{document}